\documentclass[]{emulateapj}
\usepackage{lscape}

\def\gsim{\mathrel{\hbox{\rlap{\hbox{\lower4pt\hbox{$\sim$}}}\hbox{$>$}}}}
\def\lsim{\mathrel{\hbox{\rlap{\hbox{\lower4pt\hbox{$\sim$}}}\hbox{$<$}}}}

\def\ax{{\it AEGIS-X}}
\def\chandra{{\it Chandra}}
\def\xmm{{\it XMM-Newton}}

\slugcomment{Accepted for publication in ApJ Supp}

\shorttitle{AEGIS-X}
\shortauthors{E. S. Laird et al.}

\begin{document}


\title{{\it AEGIS-X:} The Chandra Deep Survey of the Extended Groth Strip}


\author{E. S. Laird\altaffilmark{1},  K. Nandra, A. Georgakakis, J. A. Aird}
\affil{Astrophysics Group, Imperial College London, Blackett Laboratory, Prince Consort Road, London SW7 2AZ, United Kingdom}
\author{P. Barmby\altaffilmark{2}, 
C. J. Conselice\altaffilmark{3}, 
A. L. Coil\altaffilmark{4},
M. Davis\altaffilmark{5}, 
S. M. Faber\altaffilmark{6}, 
G. G. Fazio\altaffilmark{7}, 
P. Guhathakurta\altaffilmark{6},
D. C. Koo\altaffilmark{6}, 
V. Sarajedini\altaffilmark{8},
C. N. A. Willmer\altaffilmark{4}
}
\altaffiltext{1}{e.laird@imperial.ac.uk}
\altaffiltext{2}{Department of Physics and Astronomy, University of Western Ontario, London, Ontario N6A 3K7, Canada}
\altaffiltext{3}{University of Nottingham, School of Physics and Astronomy, Nottingham NG7 2RD, UK}
\altaffiltext{4}{Steward Observatory, University of Arizona, Tucson, AZ 85721}
\altaffiltext{5}{Department of Astronomy, University of California, Berkeley, CA 94720}
\altaffiltext{6}{UCO/Lick Observatory, University of California, Santa Cruz, CA 95064}
\altaffiltext{7}{Harvard-Smithsonian Center for Astrophysics, Cambridge, MA 02138}
\altaffiltext{8}{University of Florida, Department of Astronomy, Gainesville, FL 32611}


\begin{abstract}

We present the {\it AEGIS-X} survey, a series of deep \chandra\ ACIS-I observations of the Extended Groth Strip. The survey comprises pointings at 8 separate positions, each with nominal exposure 200ks, covering a total area of approximately 0.67 deg$^{2}$ in a strip of length 2 degrees. We describe in detail an updated version of our data reduction and point source detection algorithms used to analyze these data. A total of 1325 band-merged sources have been found to a Poisson probability limit of $4 \times 10^{-6}$, with limiting fluxes of $5.3 \times 10^{-17}$ erg cm$^{2}$ s$^{-1}$ in the soft (0.5--2 keV) band and $3.8 \times 10^{-16}$ erg cm$^{-2}$ s$^{-1}$ in the hard (2--10 keV) band. We present simulations verifying the validity of our source detection procedure and showing a very small, $<1.5$\%, contamination rate from spurious sources. Optical/NIR counterparts have been identified from the DEEP2, CFHTLS, and Spitzer/IRAC surveys of the same region. Using a likelihood ratio method, we find optical counterparts for 76\% of our sources, complete to $R_{AB}$=24.1, and, of the 66\% of the sources that have IRAC coverage, 94\% have a counterpart to a limit of 0.9 $\mu$Jy at 3.6 $\mu m$ ($m_{AB}$=23.8). After accounting for (small) positional offsets in the 8 \chandra\ fields, the astrometric accuracy of the \chandra\ positions is found to be $0\farcs8$ RMS, however this number depends both on the off-axis angle and the number of detected counts for a given source. All the data products described in this paper are made available via a public website.
\end{abstract}





\section{Introduction}
\label{sec:intro}

There is intense current interest in the formation of galaxies, the history of star formation and accretion power in the universe, and the inter-relations between these phenomena. This has motivated the investment of major observational resources in obtaining images and spectroscopy in various areas of the sky. These multi-wavelength surveys cover a wide range of areas, depths, angular resolutions and wavebands, ranging from very deep ``pencil" beam surveys in small areas of the sky  to wider surveys at shallower depths.

Surveys at X-ray wavelengths provide an important component of the multiwavelength arsenal, primarily because of the efficiency and sensitivity of X-ray emission in selecting Active Galactic Nuclei (AGN). There is apparently a strong relationship between AGN and their host galaxy bulges \citep[e.g.][]{fm00,ge00}, so studying the growth of supermassive black holes in the context of the evolution of their host galaxies is likely to be a fruitful area of exploration. X-ray surveys also enable an estimate of the history of accretion power in the universe and the origin of the X-ray background \citep[e.g.][]{bh05}. 

The deepest X-ray surveys in existence are the \chandra\ Deep Fields North and South (hereafter the CDF-N and CDF-S, respectively), which each cover an area of $\sim 0.1$~deg$^{2}$ to nominal depths of $\sim 2-3 \times 10^{-17}$~erg cm$^{-2}$ s$^{-1}$ \citep{al03,lu08}. The widest surveys to resolve a significant fraction of the X-ray background cover of order 10 deg$^{2}$, but to much shallower (by a factor $\sim 100$) depths (e.g. XBootes: \citealt{mu05}; \citealt{ke05}; XMM-LSS: \citealt{pi04}). The major advantage of ultradeep surveys is that they are able to probe into the distant universe, to detect ``typical" objects at high redshift. On the other hand, larger areas are required to sample significant large scale structures in the universe and hence determine the relationship between galaxy evolution and local environmental density. Moreover, larger area surveys are able to sample and find unusual, rare objects. For a complete picture it is clearly also necessary to explore the parameter space in between the ultradeep and ultrawide surveys. 

Motivated by this, we have obtained deep X-ray observations using the \chandra\ X-ray observatory in a region of sky of area $\sim 0.5$~deg$^{2}$ known as the Extended Groth Strip (EGS), covering the energy range 0.5--7~keV. Extensive multiwavelength data in this region have been obtained as part of the ``AEGIS" project \citep{da07} making it one of the premier datasets to study the co-evolution of black hole accretion and galaxy formation. The purpose of the present paper is to describe the X-ray dataset and reduction, and present a catalog of point sources derived from the \chandra\ X-ray survey, which we designate \ax. 

The structure of the paper is as follows. In \S\ref{sec:dr} we describe the data and the reduction, including the method used to calculate the \chandra\ point spread function. Detailed descriptions of our source detection and photometry procedures are given in \S\ref{sec:detphot}. In \S\ref{sec:results} the results of the source detection and analysis are presented. In \S\ref{sec:counterpart} the optical and infrared counterparts to the X-ray sources are provided and in \S\ref{sec:conc} the conclusions are given.

\section{Observations and Data Reduction}
\label{sec:dr}


The data used in this paper were obtained during two \chandra\ observation cycles. An initial \chandra\ observation in the EGS region was taken in August 2002 during Cycle 3. These observations of the ``Groth-Westphal Strip'' (GWS) region have been reported by \citet[hereafter N05]{n05}. The majority of the \ax\ data were obtained over the period March--December 2005 as part of \chandra\ Cycle 6. We analyze all the \ax\ data here in a uniform fashion. All the observations were taken using the ACIS-I instrument without any grating in place. The S2 and S3 chips of the ACIS-S array were also sometimes operating during the observations but as these are far off-axis we do not consider the data further. Details of the \chandra\ observations are shown in Table~\ref{tab:obs}.

\begin{deluxetable*}{cccccccc}
\tabletypesize{\scriptsize}
\tablecaption{Observation Log \label{tab:obs}}
\tablewidth{0pt}
\tablehead{
\colhead{Field\tablenotemark{a}} &  
\colhead{ObsID\tablenotemark{b}} & 
 \colhead{RA\tablenotemark{c}} & 
\colhead{DEC\tablenotemark{c}} & 
\colhead{Start Time\tablenotemark{d}} & 
\colhead{Mode\tablenotemark{e}} &
\colhead{On time\tablenotemark{f}} &
\colhead{Exposure\tablenotemark{g}} \\
\colhead{Name} &
\colhead{} &
\colhead{(J2000)} & 
\colhead{(J2000)} & 
\colhead{(UT)} & 
\colhead{} & 
\colhead{(ks)} & 
\colhead{(ks)} 
}
\startdata
EGS-1   & 5841   & 14 22 41.88   & +53 25 53.38   & 2005-03-14 00:04:09   & VFAINT   & 44.46   & 44.45 \\
EGS-1   & 5842   & 14 22 41.52   & +53 25 53.09   & 2005-03-16 15:54:34   & VFAINT   & 46.43   & 46.42 \\
EGS-1   & 6210   & 14 22 43.32   & +53 25 24.26   & 2005-10-03 14:56:50   & VFAINT   & 46.35   & 45.94 \\
EGS-1   & 6211   & 14 22 43.68   & +53 25 26.18   & 2005-10-12 11:43:28   & VFAINT   & 36.07   & 35.64 \\
EGS-1   & 7180   & 14 22 43.68   & +53 25 26.18   & 2005-10-13 05:16:04   & VFAINT   & 20.85   & 16.67 \\
EGS-2   & 5843   & 14 21 32.04   & +53 13 42.86   & 2005-03-19 17:13:09   & VFAINT   & 44.47   & 44.46 \\
EGS-2   & 5844   & 14 21 31.68   & +53 13 42.52   & 2005-03-21 22:37:40   & VFAINT   & 45.85   & 45.85 \\
EGS-2   & 6212   & 14 21 33.84   & +53 13 14.74   & 2005-10-04 22:56:06   & VFAINT   & 46.29   & 46.28 \\
EGS-2   & 6213   & 14 21 33.84   & +53 13 15.02   & 2005-10-06 06:52:13   & VFAINT   & 47.93   & 47.46 \\
EGS-3   & 5845   & 14 20 27.24   & +53 02 15.70   & 2005-03-24 14:33:31   & VFAINT   & 48.41   & 48.40 \\
EGS-3   & 5846   & 14 20 26.88   & +53 02 15.20   & 2005-03-27 04:51:15   & VFAINT   & 49.41   & 49.40 \\
EGS-3   & 6214   & 14 20 29.04   & +53 01 47.16   & 2005-09-28 08:09:03   & VFAINT   & 47.93   & 47.50 \\
EGS-3   & 6215   & 14 20 29.04   & +53 01 47.35   & 2005-09-29 15:58:09   & VFAINT   & 50.30   & 48.63 \\
EGS-4   & 5847   & 14 19 22.44   & +52 50 44.27   & 2005-04-06 20:01:09   & VFAINT   & 44.99   & 44.52 \\
EGS-4   & 5848   & 14 19 22.44   & +52 50 44.26   & 2005-04-07 21:03:59   & VFAINT   & 44.46   & 44.40 \\
EGS-4   & 6216   & 14 19 24.60   & +52 50 17.38   & 2005-09-20 09:35:13   & VFAINT   & 49.92   & 49.48 \\
EGS-4   & 6217   & 14 19 24.60   & +52 50 17.39   & 2005-09-23 01:34:59   & VFAINT   & 49.92   & 49.45 \\
EGS-5   & 5849   & 14 18 21.60   & +52 38 50.53   & 2005-10-11 12:47:43   & VFAINT   & 49.92   & 49.46 \\
EGS-5   & 5850   & 14 18 21.60   & +52 38 51.16   & 2005-10-14 05:15:37   & VFAINT   & 45.95   & 40.42 \\
EGS-5   & 6218   & 14 18 21.60   & +52 38 49.54   & 2005-10-07 05:31:36   & VFAINT   & 40.90   & 40.58 \\
EGS-5   & 6219   & 14 18 21.24   & +52 38 47.33   & 2005-09-25 15:57:04   & VFAINT   & 49.90   & 49.43 \\
EGS-6   & 5851   & 14 16 26.04   & +52 19 52.37   & 2005-10-15 03:03:18   & VFAINT   & 36.08   & 35.68 \\
EGS-6   & 5852   & 14 16 26.40   & +52 20 06.21   & 2005-12-03 13:00:33   & VFAINT   & 10.62   & 10.20 \\
EGS-6   & 6220   & 14 16 24.96   & +52 19 46.68   & 2005-09-13 09:17:01   & VFAINT   & 37.90   & 34.84 \\
EGS-6   & 6221   & 14 16 24.96   & +52 19 46.81   & 2005-09-15 22:11:12   & VFAINT   & 4.15   & 3.71 \\
EGS-6   & 6391   & 14 16 24.96   & +52 19 46.42   & 2005-09-16 20:43:01   & VFAINT   & 8.72   & 7.47 \\
EGS-6   & 7169   & 14 16 26.40   & +52 20 06.33   & 2005-12-06 02:29:46   & VFAINT   & 16.30   & 16.03 \\
EGS-6   & 7181   & 14 16 26.04   & +52 19 52.37   & 2005-10-15 21:17:21   & VFAINT   & 16.38   & 15.98 \\
EGS-6   & 7188   & 14 16 26.04   & +52 20 05.96   & 2005-12-05 04:50:40   & VFAINT   & 3.32   & 2.58 \\
EGS-6   & 7236   & 14 16 26.40   & +52 20 06.16   & 2005-11-30 19:29:34   & VFAINT   & 20.46   & 20.02 \\
EGS-6   & 7237   & 14 16 26.40   & +52 20 06.14   & 2005-12-04 05:26:20   & VFAINT   & 17.09   & 16.93 \\
EGS-6   & 7238   & 14 16 26.40   & +52 20 06.07   & 2005-12-03 10:02:10   & VFAINT   & 9.70   & 9.53 \\
EGS-6   & 7239   & 14 16 26.40   & +52 20 07.61   & 2005-12-11 08:31:06   & VFAINT   & 16.23   & 16.03 \\
EGS-7   & 5853   & 14 15 23.76   & +52 08 16.36   & 2005-10-16 20:16:24   & VFAINT   & 42.99   & 42.47 \\
EGS-7   & 5854   & 14 15 23.40   & +52 08 12.92   & 2005-09-30 23:52:23   & VFAINT   & 50.51   & 50.07 \\
EGS-7   & 6222   & 14 15 22.68   & +52 08 09.86   & 2005-08-28 17:20:24   & VFAINT   & 35.00   & 34.69 \\
EGS-7   & 6223   & 14 15 22.68   & +52 08 09.90   & 2005-08-31 05:06:47   & VFAINT   & 50.69   & 49.51 \\
EGS-7   & 6366   & 14 15 22.68   & +52 08 09.79   & 2005-09-03 06:30:11   & VFAINT   & 14.59   & 13.88 \\
EGS-7   & 7187   & 14 15 23.76   & +52 08 16.26   & 2005-10-17 19:07:08   & VFAINT   & 6.70   & 6.59 \\
EGS-8   & 3305   & 14 17 43.08   & +52 28 25.21   & 2002-08-11 21:43:57   & FAINT   & 29.41   & 29.36 \\
EGS-8   & 4357   & 14 17 43.08   & +52 28 25.21   & 2002-08-12 22:32:00   & FAINT   & 86.12   & 84.36 \\
EGS-8   & 4365   & 14 17 43.08   & +52 28 25.21   & 2002-08-21 10:56:53   & FAINT   & 84.22   & 83.75 
\enddata
\tablenotetext{a}{Field Name: note that EGS-8 field is the original ``Groth-Westphal Strip" described by N05}
\tablenotetext{b}{\chandra\ Observation ID}
\tablenotetext{c}{Nominal position of pointing (J2000)}
\tablenotetext{d}{Start date and time (UT)}
\tablenotetext{e}{Observation mode}
\tablenotetext{f}{Raw exposure time}
\tablenotetext{g}{Exposure time after data screening as described in \S\ref{sec:reduction}}
\end{deluxetable*}

\begin{figure*}
\begin{center}
\epsscale{1.0}
{\scalebox{1.0}
{\includegraphics{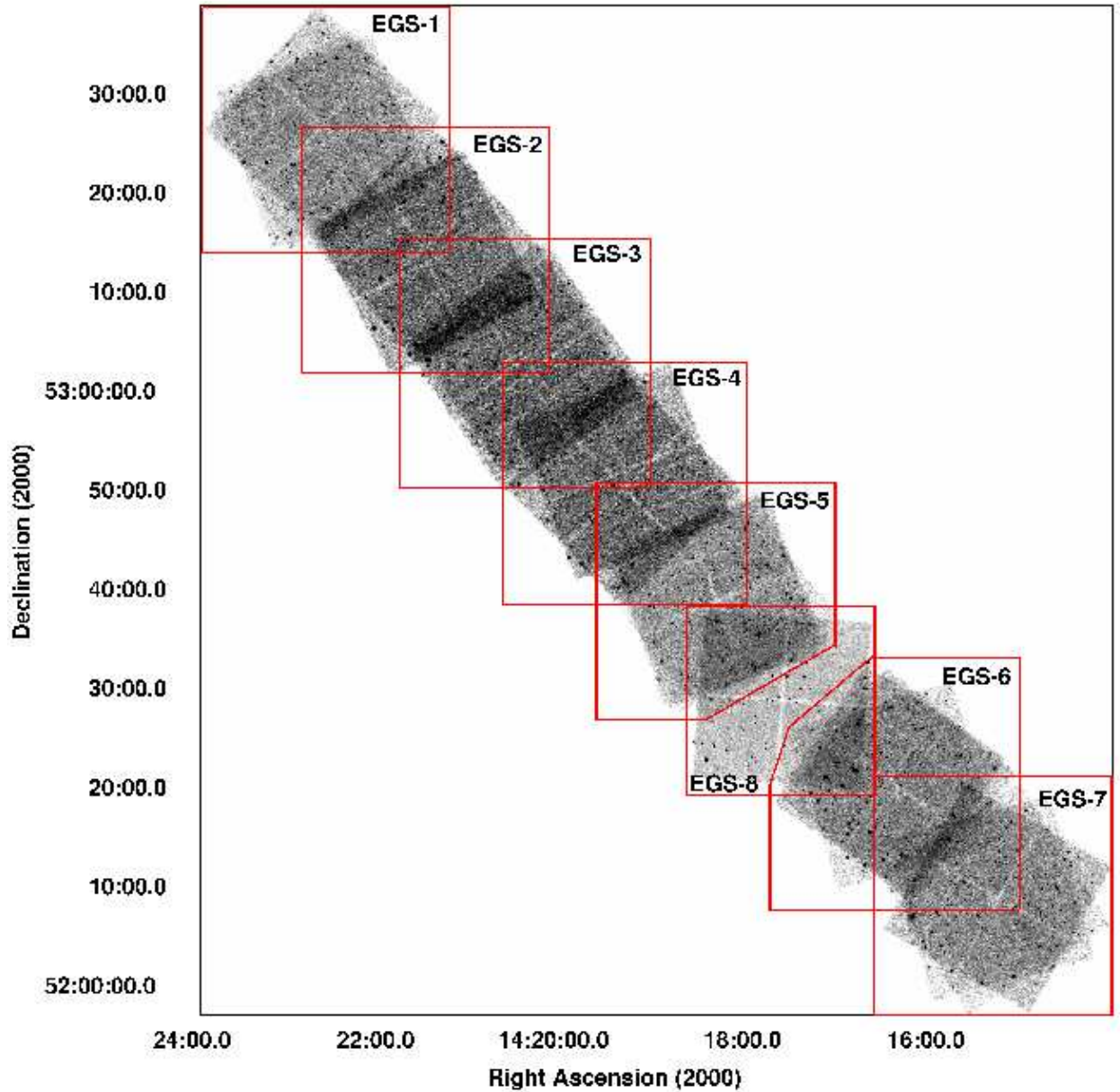}
}}
\caption{Mosaic full band image of the Extended Groth Strip showing the location and overlap of the 8 analysis fields. The EGS-8 field is the original ``Groth-Westphal Strip'' data described by N05.   
\label{fig:mosaic}}
\end{center}
\end{figure*}







\subsection{Data reduction}
\label{sec:reduction}
The data reduction was performed using the CIAO data analysis software version 3.3. Basic data reduction proceeded in a manner similar to that described in N05. 

Initially, each observation ID (obsID) was analyzed separately. We first corrected the raw (level 1) event files for any known aspect offsets. Hot pixels and cosmic ray afterglows were removed using the CIAO {\tt acis\_run\_hotpix} task for EGS observations 1--7, which were taken in {\sc vfaint} mode. Application of this task to the EGS-8 data (i.e. the original GWS data), which were taken in {\sc faint} mode, resulted in images with a large number of apparently spurious sources, which we identified post hoc due to the anomalously low optical/IRAC identification rate in this field. Application of the older afterglow removal tool {\tt acis\_detect\_afterglow} yielded source numbers and identification rate similar to the other fields. This suggests that the newer CIAO algorithm fails to identify some afterglows and/or hot pixels in the {\sc faint} mode data. We therefore adopted the {\tt acis\_detect\_afterglow} algorithm for the EGS-8 field. 

To produce a level 2 (screened) event file we applied the charge-transfer inefficiency (CTI) and time-dependent gain correction, removed the ACIS pixel randomization and applied PHA randomization, as recommended. In the case of {\sc vfaint} mode data (EGS-1--7) we also applied the ACIS particle background cleaning algorithm.

To identify periods of anomalously high background, which would hamper efficient source detection, we created light curves in the 0.5--7 keV band (excluding sources detected using the {\tt celldetect} algorithm) with a bin size of 50s. Background flares were identified using  the procedure described in \citet{n07}, whereby a quiescent background level is determined by calculating the count rate limit at which the excess variance of the background light curve is equal to zero. For our \chandra\ ACIS data, we excluded times when the background count rate exceeded 1.4 times this limit (c.f. Nandra et al. 2007, who used two times the limit for \xmm\ EPIC-pn background cleaning). This procedure was found to be adequate in most cases but some residual background flares not identified by this method were removed manually. One observation, obsID 4365 in field EGS-8, exhibited a $\sim$25ks period of elevated, but relatively stable and well-behaved, background which would be excluded using the filtering criteria described above. As described in N05, the inclusion of this $\sim$25ks increases the sensitivity of the observation to point sources and therefore it has not been filtered from the data set for this work. 

From the cleaned level 2 event file we created images in four energy bands, which we designate full band (FB; 0.5--7 keV), soft band (SB; 0.5--2 keV), hard band (HB: 2--7 keV), and ultrahard band (UB; 4--7 keV). We used the standard CIAO procedure ({\tt merge\_all}) to produce exposure maps for each obsID. These account fully for the telescope and instrument efficiencies, as well as the chip gaps and dithering during the observation. The efficiencies are also a function of photon energy, so creating the exposure maps requires an assumption about the energy distribution of the detected photons, i.e. the source spectrum. N05 calculated these exposure maps at a single energy representative of the detected photons in each band. Following this procedure through to the source photometry stage indicated that these exposure maps produced by this method resulted in hard band fluxes that were systematically low. We verified this using MARX simulations (see \S\ref{sec:astrom}), inputting synthetic sources of known flux and running through the entire source detection and photometry procedure. Small systematic flux underestimations were also found for the other bands, but the HB suffered the most severe problem, at the $\sim 20$\% level. In part, this may be due to an inconsistency between the single energy exposure maps, and the $\Gamma=1.4$ spectrum used to convert counts to flux (see \S\ref{sec:photom}). Switching to a weighted exposure map based on a $\Gamma=1.4$ spectrum solved the flux problem, and we adopted this for the calculation of all exposure maps. 

As already discussed, and listed in Table~\ref{tab:obs}, each field typically comprises several obsIDs and, furthermore, the obsIDs from any given field can overlap the others. It is impractical to perform the analysis for the whole EGS field, as the resulting images would be too large when made at native 0.5 arcsec resolution. We therefore defined a total of 8 regions (Figure~\ref{fig:mosaic}), one for each field, which we used to create images and perform source detection, eventually merging the source catalogs from these regions. 

The field regions were defined based on the limits in sky co-ordinates (X, Y) of the first obsID with a boundary of 20 pixels in all directions. For example, for EGS-1 the field region is defined by the maximum and minimum sky co-ordinates for obsID 5841, plus or minus 20 pixels in both X and Y directions. We then identified all other obsIDs which overlap with this field box, regardless of whether they are nominally part of that field (e.g. we would combine overlapping data from EGS-2 with EGS-1 where it exists). The exception to this is that we did not combine overlapping data with the three observations that comprise EGS-8 with the adjoining fields (EGS-5 and EGS-6) due to the unfavorable combination of small and large off axis angle data and therefore data with significantly different point-spread function sizes. Source detection tests performed when combining data in this way resulted in fewer detections compared to considering the EGS-8 data by itself. For each overlapping obsID identified we registered the co-ordinates relative to the first observation in the stack using the {\tt acis\_align\_events} task. This performs a source detection using the CIAO {\tt celldetect} algorithm, at a 3$\sigma$ limit, and uses the centroids to the relatively bright sources so identified to realign the sky co-ordinates of the images. We only performed this realignment in cases where the algorithm detects at least four such sources.   Following the realignment we created a merged event file covering the defined field region, and images in the soft, full, hard, and ultrahard bands. 

\begin{figure}
\epsscale{1.0}
{\scalebox{0.675}
{\includegraphics{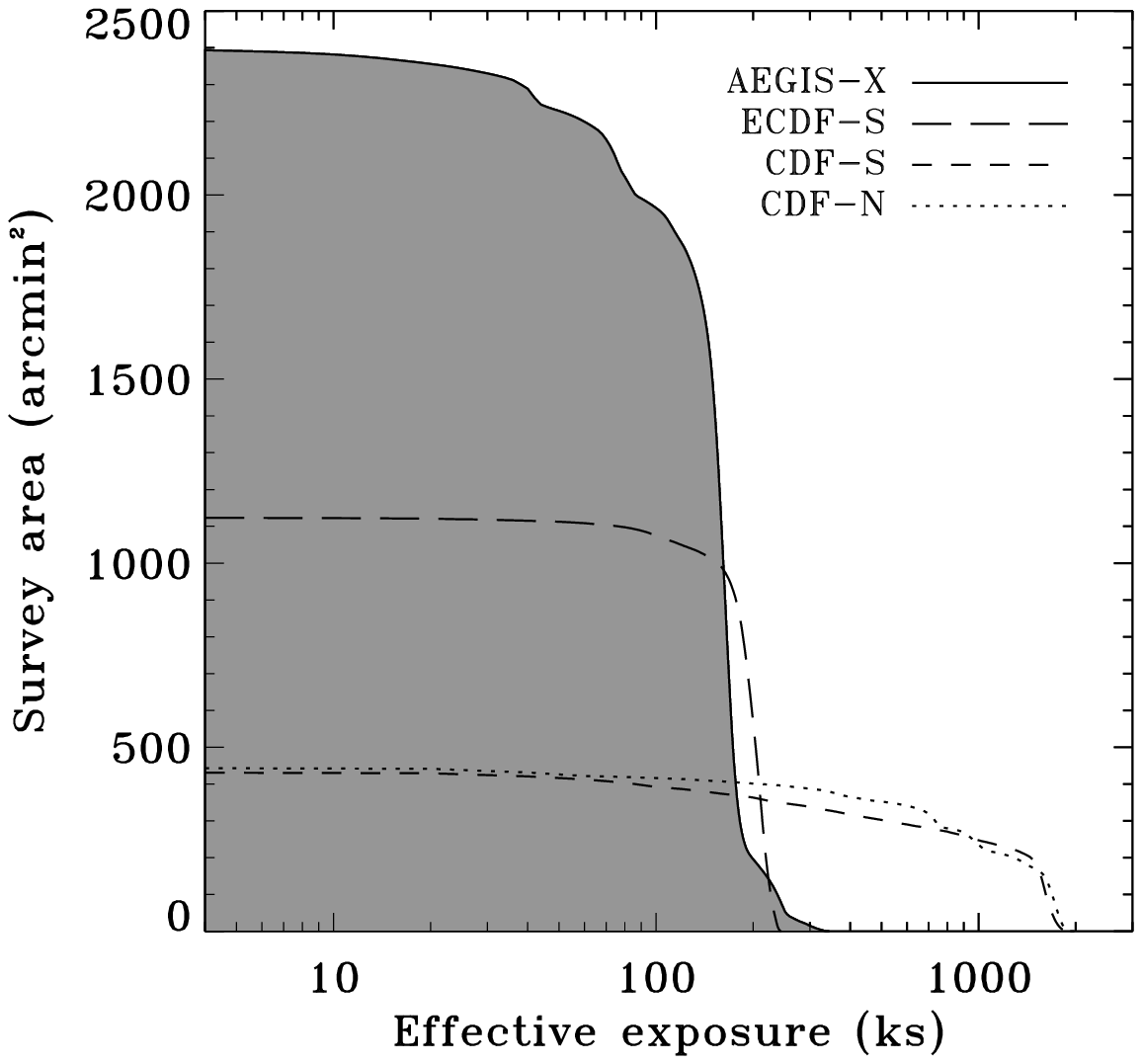}
}}
\caption{Effective exposure time versus survey area for \ax, the \chandra\ Deep Fields North and South and the Extended \chandra\ Deep Field South. These were calculated by taking the value of the full band exposure map at each pixel and dividing by the on-axis effective area. The CDFs reach much deeper exposure but \ax\ provides a large increase in area compared to these surveys at moderate depth. 
}
\label{fig:effexp}
\end{figure}

An exposure map for the field region was created by summing the exposure maps of the individual obsIDs contributing to the stack, which had been created previously. A plot of the effective exposure versus survey area is shown in Figure~\ref{fig:effexp}, comparing \ax\ to the CDFs. This shows the large increase in area for moderately deep exposures (100--200ks) afforded by our survey.

The final products of this basic data reduction are therefore the 8 event files corresponding to each field EGS-1 to EGS-8, and the corresponding images and exposure maps in the four analysis bands. These images are of manageable size to perform source detection. They also overlap, which help avoid edge effects in the source detection, but means we must subsequently delete duplicate sources detected in more than one nominal field. 

\subsection{Point Spread Functions}
\label{sec:psf}

To determine accurately the significance and flux of our sources requires a good estimate of the Point Spread Function (PSF). In the case of \chandra\ ACIS data this depends on the position on the detector and the energy. Finding previous methods for estimating the \chandra\ PSF for use in surveys to be unsatisfactory, either because they are not sufficiently accurate (in the case of the CIAO task {\tt mkpsf}) or because they are prohibitively slow (in the case of ChaRT, the \chandra\ ray-tracer simulator) we decided to develop our own procedure for calculating the PSF. The most important difference in the analysis of the data in this paper compared to that described in N05 is the use of the new calculations of the PSF. 

N05 used the CIAO task {\tt mkpsf} to estimate the PSF, which calculates a PSF image based on a look-up table based on calculations using the \chandra\ Ray Tracer PSF simulator ChaRT \citep{ca03}. ChaRT simulates photons passing through the best available model of the High Resolution Mirror Assembly (HRMA), based on extensive pre-launch testing \citep{je95}, which includes details of the support structures, stray light baffles, and an independent model of each of the mirror segments, and thus provides the best modelling of the PSF for any energy or position in the focal (detector) plane. The available ChaRT-generated library files for {\tt mkpsf} cover a relatively small number of positions and energies, and the tool must therefore interpolate to generate the necessary PSFs. In addition, for large off-axis angles the PSF images created by {\tt mkpsf} have the edges clipped, as the PSF can become larger than the generated image.  

We have therefore taken an alternative approach, and calculated the PSF using the MARX simulator \citep{wi03}. MARX is the detector simulator, but is also able to simulate photons passing through a simplified version of the ChaRT HRMA model. In particular, the MARX HRMA model does not include the physical support structure and thus PSFs generated by MARX lack some of the complicated substructure seen in the ChaRT PSFs. However, knowledge of the detailed structure within the PSF is not needed for our point source detection and photometry methods. We only require an aperture which contains a particular Enclosed Energy Fraction (EEF), and thus the simplified mirror model is sufficient for our needs, as well as requiring less computer processing time.

We have generated look-up tables of the PSF for a range of EEFs over a fine, evenly spaced grid (10 pixel spacing) covering the ACIS-I detector. At each position, an image of the PSF was generated with MARX for a monochromatic source initially of energy 1keV (representative of our soft band), using 200,000 input rays. The effects of quantum efficiency and the filter transmissions were turned off, as these properties do not affect the PSF, thus maximizing the number of detected photons for a given number of input rays, improving the counting statistics in the PSF image without requiring additional ray traces and longer processing time. At each position, counts were extracted within elliptical apertures (found to best describe the overall shape of the PSF) with increasing semi-major axes. For each value of the semi-major axis, we calculated the semi-minor axis and orientation angle of the ellipse from the moments of all counts within a circle of radius equal to the semi-major axis. We calculated the EEF for each of the set of elliptical apertures, and used interpolation to calculate the ellipse parameters for an aperture containing a particular EEF (specifically 50, 60, 70, 80, 90, and 95\%). The total counts (for normalization of the EEF) were taken from the entire area of the detector. When sources fell close to the edge of the detector chips or the chip gaps, MARX was altered to shift the nominal position of the detector in the focal plane by around 100 pixels so the PSF was fully sampled. The ellipse parameters for the desired EEFs were stored in look-up tables, providing a very fast method of determining the PSF, using the closest grid position. The PSF varies little on our sampling scale of 10 pixels, so we do not need to interpolate. The procedure was repeated using representative monochromatic energies for each of our energy bands (FB: 2.5 keV, HB: 4 keV, UB: 5.5keV).

The PSF is defined at a position on the detector (fixed relative to the mirror). However, our X-ray data are the result of a number of observations (obsIDs) with different orientations and pointing directions, merged to create images with the maximum possible exposure.  The source detection and photometry described below requires extraction of counts using a single, circular aperture from stacked images which have different PSFs. We therefore calculated exposure weighted PSFs for each candidate source in the merged images. The position in each of the constituent obsIDs was found, and in each obsID the average value of the exposure map at that position (extracted over approximately 80 surrounding pixels) was calculated, and the PSF was retrieved from the MARX table. Each of the MARX PSFs were converted from ellipses to circles with a radius equal to the square root of the product of the major and minor ellipse axes. For each candidate source the PSFs were then combined, weighting each by the ratio of exposure in a particular obsID to the exposure in the merged image. This was repeated
for each energy band.

\section{Source detection and photometry}
\label{sec:detphot}

The point-source catalog was created using an updated version of the source detection algorithm described by N05 and \citet{la06}. Briefly, for the \ax\ data, candidate sources were initially identified in each individual field in a number of different energy ranges, their significance computed, and a threshold applied. The sources considered significant were then band-merged, and photometry performed to estimate the fluxes in several energy ranges. Finally the source catalogs for the individual (overlapping) pointings were merged to remove duplicate sources for the final catalog. These steps are described in detail below. 

\subsection{Source detection}
\label{sec:srcd}

The detection algorithm described in N05 involves pre-detection at a low significance threshold followed by aperture extraction of the counts to determine the source significance. The pre-detection is performed using the CIAO wavelet-detection algorithm {\tt wavdetect}, run on the unbinned, stacked images, at wavelet scales of 1, $\sqrt{2}$, 2, 2$\sqrt{2}$, 4, 4$\sqrt{2}$, 8, 8$\sqrt{2}$, and 16 pixels. Only positions with at least 10\% of the maximum field exposure were considered, and the procedure is run at a significance threshold of $10^{-4}$ on images in all four bands. The minimum exposure requirement means that in practise our source catalog will not cover the full 0.64 degree$^{2}$ of the survey: the least sensitive 3.2\% of the total area is excluded. The chosen low threshold for {\tt wavdetect} is likely to result in a large fraction of spurious sources, but very few real sources should be missed. To determine the source significance we first calculated the 70\% EEF PSFs for the stacked image, by exposure weighting the individual PSFs from each individual exposure in the stack as described above. 

Using the 70\% EEF PSF (determined a priori to be the most efficient radius for source detection) we extracted the total counts for each candidate source, and the effective exposure, the latter being the value of the exposure map averaged over all the pixels in the extraction region. Background counts were determined by first masking the image to remove pixels within the 95\% EEF of each candidate {\tt wavdetect} source. We then extracted background counts from an annulus with an inner radius equal to the 1.5 times the 95\% EEF radius for the source, and an outer radius 100 pixels more than this. An average exposure map value was calculated for this background area also. The counts in the background area were than scaled to the source region by the ratio of the source and background areas, and the ratio of source and background average exposure. 

For each candidate source in each energy band, the Poisson probability that the total counts in the source region would be observed based on the expected background counts was then calculated. Following N05, we adopted a significance of $4 \times 10^{-6}$. Typically, when run at a significance of $10^{-4}$ {\tt wavdetect} identifies 400--500 candidate sources in each \ax\ field, but more than half of these are found to be insignificant -- and hence probably background fluctuations -- when we perform the aperture extraction and Poisson probability calculations. Because we initially masked out all the {\tt wavdetect} candidate sources we may therefore have incorrectly masked out the largest positive background fluctuations, leading to an underestimate of the true background level. For this reason we perform a second pass of the source detection, masking out only the sources with Poisson probability $<4 \times 10^{-6}$ when calculating the background. We then recompute these probabilities with the new background estimates and reapply the threshold. This second calculation of the Poisson probability is a refinement of the method described in N05 and \citet{la06}. Technically further iterations of this kind could be applied but in practice there is negligible difference in the background estimates after two iterations. 

When run at low thresholds, such as we have used for the seed catalog, {\tt wavdetect} sometimes detects the same source twice, particularly at larger off-axis angles where the counts are spread out over a wide area. Visually inspecting the images it is clear that only one source is present. During the source detection phase we therefore check for sources that are separated by less than 5 pixels. If there are any then we remove from the list the source position with fewer counts in the 70\% EEF area. There are a total of 4 such cases in \ax. 

\subsection{Band-merging}
\label{sec:bandmerge}

Thus far the source detection procedure has been carried out separately for the soft, full, hard and ultrahard band images. While it is possible that a source will be detected in just one of these bands, in practice the majority of course are detected in several, necessitating a band-merging procedure to produce a single source catalog for that field. In practice most objects are detected in the full band, so we matched the soft, hard and ultrahard band source lists with the full band. The positional accuracy of \chandra\ depends on the off-axis angle (OAA), due the position dependence of the PSF, so we adopted a variable radius for source matching. The cross-matching tolerances are shown in Table~\ref{tab:xmatch}, and are based on positional accuracies as determined in \S\ref{sec:astrom}. We used a 3$\sigma$ radius, combining the uncertainty in the two positions in quadrature. If more than one cross-band counterpart was identified within the tolerance the closer match was selected. There was only one case of this in the survey, in EGS-5. Two soft band sources are detected within the match tolerance radius from a single full band source. Visual inspection of this apparent ``double"  suggested that it was in fact a single object detected twice with a slightly different position. The soft position with the larger offset from the full band source was therefore removed from the catalog.

The source positions used in the band merged catalog are decided as follows: if a significant full band counterpart exists then the full band position is used. If not and a soft band counterpart exists then
the soft band position is used. The hard band position is used for hard band sources without full or soft counterparts. Ultrahard band positions are used for sources only detected in this band. 

After performing the band merging we visually inspected the images and source list in each of the fields and checked that the correct cross-band counterparts were identified. 

\begin{deluxetable}{ccc}
\tabletypesize{\small}
\tablecaption{Cross-band matching radii \label{tab:xmatch}}
\tablewidth{0pt}
\tablehead{
\colhead{OAA\tablenotemark{a}} &   \colhead{$r_{1 \sigma}$\tablenotemark{b}} &  \colhead{Tolerance\tablenotemark{c}} \\
\colhead{(arcmin)} &  \colhead{(arcsec)} & \colhead{(arcsec)} 
}
\startdata
0--3 & 0.30 & 1.30 \\
3--6 & 0.57 & 2.44 \\
6--9 & 1.13 & 4.79 \\
$>9$ & 1.67 & 7.08 
\enddata
\tablenotetext{a}{Off-axis angle}
\tablenotetext{b}{$1\sigma$ positional uncertainty determined from MARX simulations (see text)}
\tablenotetext{c}{Cross-band match tolerance}
\end{deluxetable}

\subsection{Photometry}
\label{sec:photom}

As with source detection, photometry was carried out on each of the eight individual EGS fields. For each source in the band-merged catalog we extracted total counts and effective exposure values from the 90\% EEF area in each analysis band. Background counts and exposure values were then extracted in these same bands from an image in which the 95\% EEF area of all the significantly detected sources (regardless of band) had been masked out. The background extraction region was defined in the same way as it was for the detection (\S\ref{sec:srcd}). Background counts were once more scaled based on the ratio of the areas of the source and background cells and the ratio of the average effective exposures in all pixels in the cells. These scaled background counts were then subtracted from the total counts to give the net source count.  

To identify sources that were close to each other, and hence possibly have confused photometry, we identified those that were separated by more than the match tolerances given in Table~\ref{tab:xmatch} but whose $90$\% EEF radii overlap. These are flagged in the catalog as having photometry that is likely contaminated by a nearby object. In some cases, the actual position can fall within the 90\% EEF radius of the nearby source. In these cases the photometry will be heavily contaminated by the nearby source, and they are flagged as being confused. In such cases the positions may also be highly inaccurate. 

We calculated source fluxes and 1$\sigma$ confidence limits from the detected counts using a Bayesian method that corrects for the Eddington bias and is based on \citet{kr91}. 
This methodology assumes the range of possible source fluxes is a continuous distribution fully described by the observed data and prior knowledge of the physical system. We apply Bayes' theorem,
\begin{equation}
P(S | N, B) \propto {\cal L} (N | S,B) \pi(S)
\label{eq:postdist}
\end{equation}
where $P(S|N,B)$ is the posterior distribution function for the source rate, $S$.
${\cal L}(N | S,B)$ is the Poisson likelihood of obtaining the observed data, $N$ counts, for a given source rate, $S$, and background $B$. Thus,
\begin{equation}
{\cal L}(N|S,B) = \frac{(S+B)^N}{N!}e^{-(S+B)}
\end{equation}
This approach naturally incorporates the contribution of both source and background components to the total observed counts, and thus provides a better description of the nature of faint sources than classical approaches normally employed \citep[e.g.][]{gi02,al03,ki07}. $\pi(S)$ is the prior probability distribution, and reflects our prior knowledge of the range of possible source fluxes. \citet{kr91} adopted a constant uninformative prior, requiring only that the source flux must be greater than 0. However, the distribution of fluxes of X-ray sources (the log$N - $log$S$) are found to be well described by a broken power-law \citep{ge08}. This introduces Eddington bias \citep{ed40} which affects the classical estimate of source fluxes since a larger number of faint sources are scattered to higher fluxes than vice-versa. Thus the fluxes of faint sources are generally over-estimated. We can correct for the Eddington bias with our Bayesian method, by adopting a prior based on the observed log$N - $log$S$ relation. To approximate the log$N - $log$S$ we use a faint end slope of -1.5 and a bright end slope of -2.5 for each of the analysis bands. We use break fluxes for each band as given in \citet[Table 2]{ge08}. The best estimate of the source flux is then obtained by finding the mode of the posterior distribution. Differentiating equation \ref{eq:postdist} provides an analytic solution for a prior with power-law slope $\beta$, 
\begin{equation}
\hat S = {\textstyle\frac{1}{2}} \left( N - B + \beta + \sqrt{ (N-B+\beta)^2 + 4B\beta }\right)
\end{equation}
This expression is easily extended to the case of a broken power-law. 
An additional condition for the existence of a solution is introduced,
\begin{equation}
(N-B+\beta)^2 \geq -4B\beta
\end{equation}
When there is no solution, we only quote upper limits on the source fluxes.

For the Bayesian approach, confidence limits are obtained by integrating the posterior distribution, 
\begin{eqnarray}
\mathrm{CL} &=&  \int^{S_u}_{S_l} {\cal L} ( N | S, B) \pi(S) \;\mathrm{d}S  \\
                        &=&  \int^{S_u}_{S_l} K \frac{(S + b)^N}{N!} e^{-(S+b)} \pi(S) \;\mathrm{d}S\nonumber
\end{eqnarray}
where CL is a given confidence level, and $K$ is a normalization constant. For a faint-end slope of $\beta_1=-1.5$ the posterior distribution diverges at the faintest fluxes. Extending the power-law distribution to the faintest fluxes is clearly unphysical, and makes normalization impossible. We thus set a lower limit to the possible source fluxes which is 10 times fainter than the detection limit at the position of each source, given the background. Following \citet{kr91}, we adopt Highest Posterior Density intervals, which minimize the size of the confidence interval for the given confidence level (68\%), and thus sets the tightest limits on the source flux. These are calculated by numerical integration of equation 5. 

We use this method to calculate the fluxes of all sources in each of the four bands, regardless of their Poisson probability or net counts. While many sources are not detected in all four analysis bands, we can nonetheless often determine reasonably reliable fluxes in those undetected bands. Effective on-axis source count rates were calculated by dividing the net counts with the average value of the exposure map, and aperture correcting.  Count rates were converted to fluxes using a $\Gamma=1.4$ spectrum with Galactic $N_{\rm H}$ of  $1.3\times10^{20}$~cm$^{-2}$ \citep{dl90}. The count rates in the full, hard and ultrahard bands were also extrapolated to the standard bands: 0.5--10, 2--10, and 5--10~keV, respectively. 

\begin{figure}
\vspace{-0.25in}
\hspace{-0.15in}
\epsscale{1.0}
{\scalebox{0.9}
{\includegraphics[angle=0]{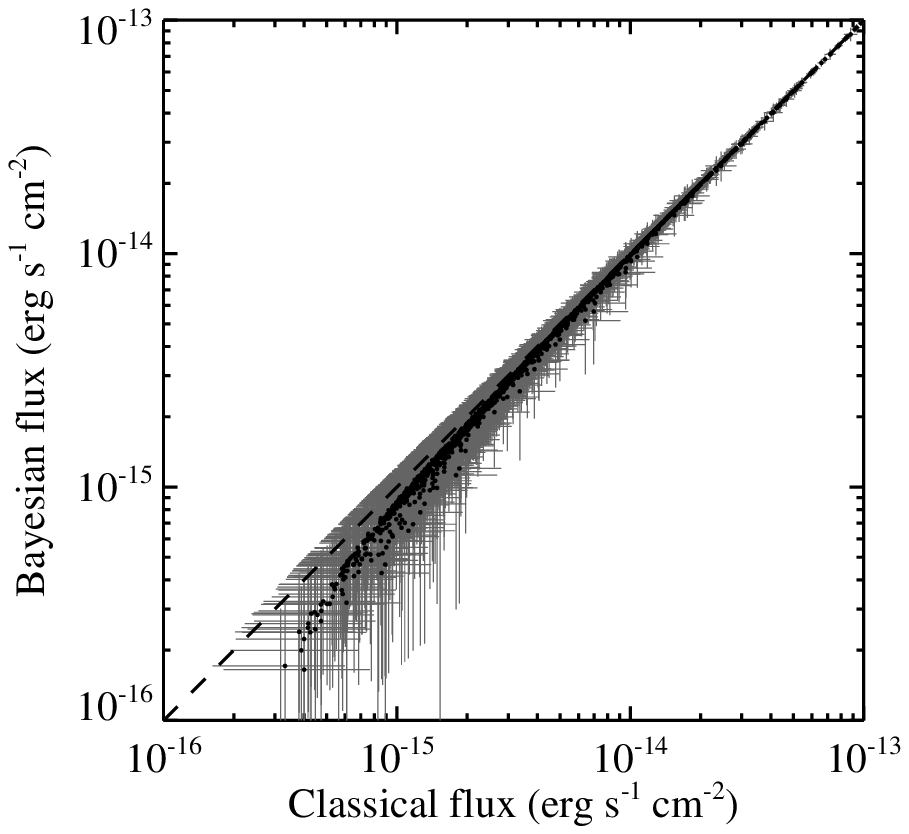}}
}
\vspace{-0.2in}
\hspace{-0.15in}
\epsscale{1.0}
{\scalebox{0.9}
{\includegraphics[angle=0]{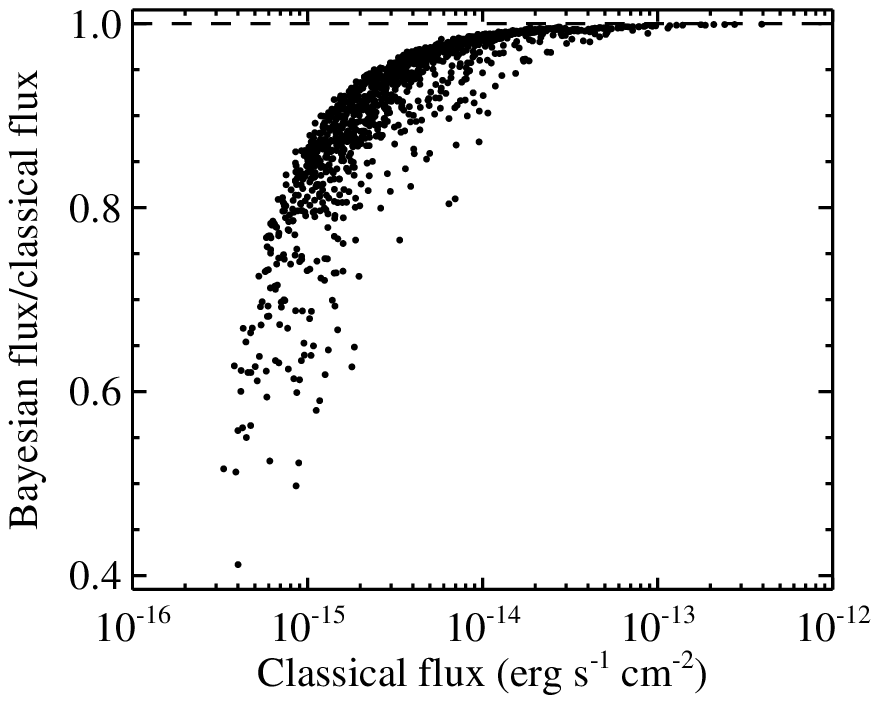}}
}
\caption{Comparison of a classical and a Bayesian method (which corrects for the Eddington bias) of calculating fluxes. (Top) Classical full band flux (0.5--10 keV) versus Bayesian flux. Errors bars show the 1$\sigma$ confidence limits for both methods. (Bottom) Classical full band flux versus the ratio of Bayesian to classical flux. At the brightest fluxes both methods show good agreement. At fainter fluxes the classical method can over predict the fluxes by up to 100\%, due to the Eddington bias. The same trend is seen in the soft, hard, and ultrahard bands.  
}
\label{fig:fluxes}
\end{figure}

For comparison, we also calculated fluxes using a classical method, with the 1$\sigma$ confidence ranges on the detected counts determined according to the prescription of \citet{ge86}. In Figure~\ref{fig:fluxes} we compare the Bayesian and classical methods for calculating fluxes, for full band sources. For bright sources, $f_{0.5-10~\mathrm{keV}} \gtrsim 10^{-14}$ erg cm$^{-2}$ s$^{-1}$, the fluxes and errors from both methods agree well. However, the fluxes of the faintest sources do not agree because the classical method fails to correct for the Eddington bias or accurately describe the errors. In such cases the classical method can over predict the fluxes by up to factor of 2.

\subsection{Hardness ratios}
\label{sec:hr}

HRs were calaculated using a Bayesian approach, following \citet{pa06}. Their method models 
the detected counts as a Poisson distribution and gives error bars and
reliable HRs for sources with both low and high counts. To calculate the HRs with the Bayesian approach we
use the {\tt BEHR}\footnote{See
http://hea-www.harvard.edu/AstroStat/BEHR/.} package (ver. 11-08-2007,
\citealt{pa06}). For this we use a flat, non-informative prior and use
the Gaussian quadrature method for sources with less than 20 net
counts in either the soft or hard bands, and the Gibbs sampler method
for all other sources. Effective area corrections to on-axis values
are input to {\tt BEHR} to allow on-axis HRs to be calculated and we
use the mean of the posterior probability distribution as the best
estimate for the HR \citep{pa06}. 

\begin{figure}
\epsscale{1.0}
\hspace{-0.4in}
{\scalebox{0.5}
{\includegraphics{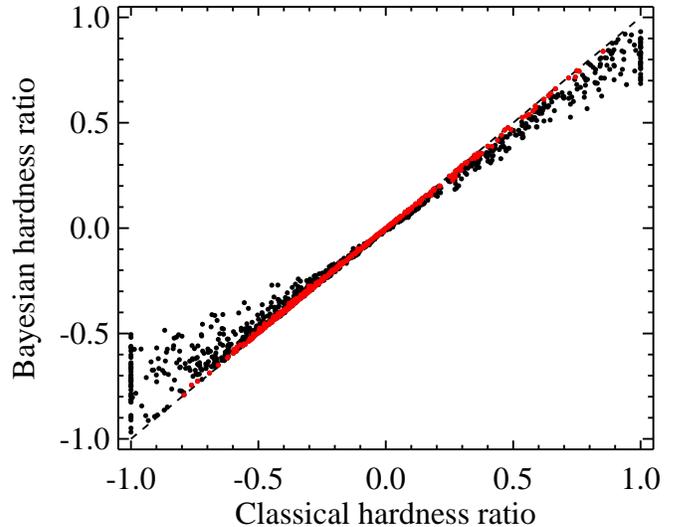}
}}
\caption{Comparison of a classical with a Bayesian method of calculating HR for all sources in the \ax\ catalog. The red points are sources with Poisson probability $p<10^{-8}$ in both soft and hard bands, for which the classical and Bayesian methods agree.}
\label{fig:HR}
\end{figure}

For comparison, we also calculated HRs using the classical method: $HR = (H-S)/(H+S)$, where $H$ and $S$ are simply hard and soft band net counts respectively, corrected to on-axis values. Where a source has only upper limits in the soft or hard bands, the HR is designated as $+1$ and $-1$, respectively. Figure~\ref{fig:HR} compares the two methods of calculating HR. There is good agreement for the bright, highly significant sources, but for weak sources the two methods disagree, particularly at low and high HR values, confirming the results of \citet{pa06} and \citet{ki07}.

\subsection{Sensitivity maps}
\label{sec:sens}

For a number  of applications, one needs to  know the flux sensitivity
of the  images as  a function  of position and  energy band.   We have
calculated  sensitivity maps,  and hence  also  the area  of the  \ax\
survey as a function of flux, using the  method described 
by \citet{ge08}.  This  accurately  estimates  the  probability  of
detecting   a  source  with   a  given   X-ray  flux   accounting  for
observational  effects, flux  estimation  biases, and  the fraction  of
spurious sources  expected in any  source catalog.  The  backbone of
the  method  is  the  Poisson probability  distribution.   For  a
detection cell with local background, $B$, one can estimate the minimum
integer number of total counts, $L$, for which the Poisson probability
that fluctuations of the background produce at least $L$ counts is less
than  the adopted  threshold for  source  detection, $4\times10^{-6}$.
Repeating this exercise  for different cells across the  image one can
determine $L$ as function of  position on the detector.  The 2-D image
of $L$ is the sensitivity map.  The sensitivity map determined in this
way  does not require  any assumptions  on the  spectral shape  of the
source.  A useful 1-D representation of
the sensitivity map  is the total detector area in  which a source with
flux  $f_X$  can be  detected.  
The sensitivity curve is estimated by adopting a
Bayesian approach.  For a source  with flux $f_X$ and a given spectral
shape ($\Gamma =  1.4$ in this paper) we  determine the probability of
detection  in a  cell with  mean  background $B$  and detection  limit
$L$. The total observed counts in the  cell are $T = B + S$, where $S$
is the mean expected source contribution.  In practise this depends on
the  observation exposure  time, the  vignetting of  the field  at the
position of  the cell, and the  fraction of the total  source counts in
the cell  because of the PSF  size.  The probability  that $T$ exceeds
$L$ is given by the cumulative Poisson probability.

\begin{figure}
\epsscale{1.0}
{\scalebox{0.775}
{\includegraphics{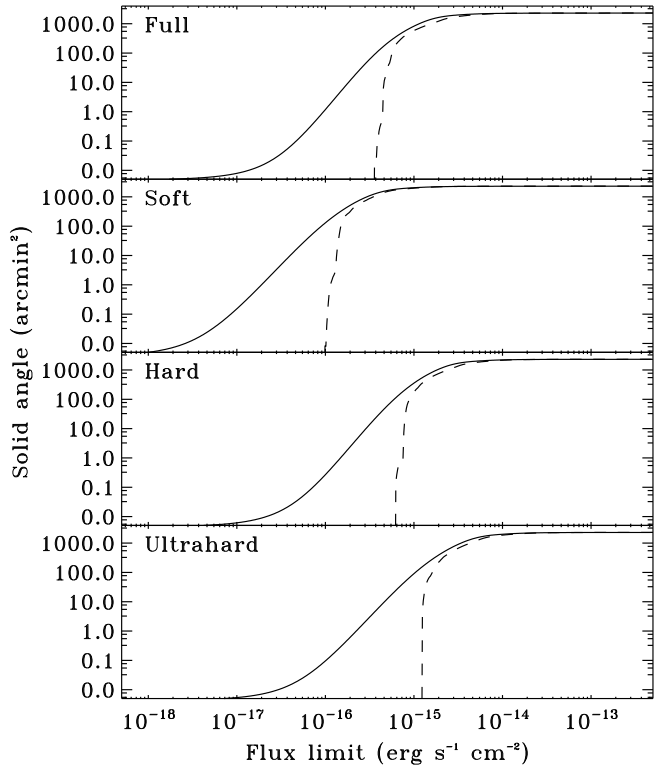}
}}
\caption{Sensitivity curves in the full, soft, hard, and ultrahard bands. The solid lines are the area curves using the Bayesian method of \citet{ge08} described in \S\ref{sec:sens}. The dashed line shows the classical area curves that do not account for the Poisson probability density distribution of the total observed counts in the detection cell.}
\label{fig:area_curve}
\end{figure}

Again, for comparison purposes, we have also calculated the sensitivity maps using
a ``classical''  method (as, e.g. N05), whereby we assign a  single limiting flux
to each detection cell.  
The sensitivity  curves in different energy bands  estimated using the
\citet{ge08} and classical methods are shown  in
Figure~\ref{fig:area_curve}.  In the classical method there is a
limiting flux below which the  sensitivity drops to zero. In contrast,
in the \citet{ge08}  method, at faint fluxes as  $S \rightarrow 0$ the
Poisson probability  converges to $p_{thresh}=4\times10^{-6}$, because
this  is the  finite probability  of a  random  background fluctuation
above the limit $L$.

\section{Point Source Catalog}
\label{sec:results}

To produce the final catalog we need both to remove duplicate sources detected in more than one of the (overlapping) fields, and to register all the images to the same astrometric frame. To do so, we first matched the X-ray catalogs to the DEEP2 optical photometry catalog \citep{co04} and the median RA and DEC offsets were calculated. The X-ray positions were shifted to be in agreement with the DEEP2 images \citep[i.e. the SDSS frame;][]{co04} using the offsets given in Table~\ref{tab:off}. 

\begin{deluxetable}{ccc}
\tabletypesize{\small}
\tablewidth{0pt}
\tablecaption{Positional offsets applied to individual fields \label{tab:off}}
\tablehead{
\colhead{Field} &   \colhead{$\Delta$RA} &  \colhead{$\Delta$DEC} \\
\colhead{} &  \colhead{(arcsec)} & \colhead{(arcsec)} 
}
\startdata
EGS-1  & -0.081 &~0.046 \\
EGS-2  & -0.031 &~0.060  \\
EGS-3  & ~0.025 &~0.048 \\
EGS-4  & -0.151 &~0.086 \\
EGS-5  & -0.197 &-0.004 \\
EGS-6  & -0.068 &-0.051  \\
EGS-7  & -0.102 &-0.002 \\
EGS-8  & -0.072 &~0.059 
\enddata
\end{deluxetable}

The source lists were then combined by finding the closest source within a match radius of $5\arcsec$. Typically the offsets were much smaller than this (see \S\ref{sec:astrom}), however as sources are matched only {\it between} individual fields and not within them there is no danger of close sources being erroneously combined. We visually compared the combined source list to the individual field source lists to confirm this. Where a source is detected in more than one field, we quote the source properties (e.g counts, probabilities, fluxes) for the field in which the source has the smaller OAA relative to the field center. Usually the source properties will be very similar in the two overlapping frames in which it is detected, but can differ subtly as e.g. the backgrounds will in general be different.  The original detected field can be identified by the `Field\_ID' tag in the source catalog. The final catalog was ordered and numbered according to increasing RA. 

\begin{deluxetable}{ccccccc}
\tabletypesize{\small}
\tablewidth{0pt}
\tablecaption{Number of sources detected using different methods\label{tab:det_methods}}
\tablehead{
\colhead{Method} & 
\colhead{$p_{thresh}$} & 
\colhead{FB\tablenotemark{a}} & 
\colhead{SB\tablenotemark{a}} & 
\colhead{HB\tablenotemark{a}} &
\colhead{UB\tablenotemark{a}} &
\colhead{Merged} 
}
\startdata
This work & $4 \times 10^{-6}$ &1221 & 1032 & 741  & 350 & 1325  \\
{\tt wavdetect} & $1 \times 10^{-7}$ &1160 & 968  & 696  & 311 & 1258  
\enddata
\tablenotetext{a}{FB = full band (0.5--7 keV), SB = soft band (0.5--2 keV), HB = hard band (2--7 keV) and UB = ultrahard band (4--7 keV).}
\end{deluxetable}

\begin{deluxetable}{lcccc}
\tabletypesize{\small}
\tablewidth{0pt}
\tablecaption{Sources detected in one band but not another\label{tab:det_sources}}
\tablehead{
\colhead{Detection} & 
\multicolumn{4}{c}{Non-detection band} \\
\colhead{band (keV)} & 
\colhead{Full} & 
\colhead{Soft} & 
\colhead{Hard} &
\colhead{Ultrahard}
}
\startdata
Full (0.5--7)   & \ldots & 274    & 497    & 876 \\
Soft (0.5--2)   & 85     & \ldots & 466    & 748 \\
Hard (2--7 )    & 17     & 175    & \ldots & 396 \\
Ultrahard (4--7)& 5      & 66     & 5   &\ldots 
\enddata
\end{deluxetable}

Once double sources are removed there are a total of 1325 distinct sources in the final catalog (Table~\ref{tab:cat1}). Details of how many sources are detected in each individual band are given in Table~\ref{tab:det_methods} and a summary of the number of sources detected in one band but not another is given in Table~\ref{tab:det_sources}. A large number of sources are detected (at $p <4 \times 10^{-6}$) in just one band, specifically, 115, 85, 14, and 2 in the full, soft, hard, and ultrahard bands respectively. The brightest significant source in the catalog has a total of 4290.5 net counts in the FB and the faintest has 4.2 net counts.

\subsection{Flux limit}

\begin{figure}
\epsscale{1.0}
{\scalebox{0.775}
{\includegraphics{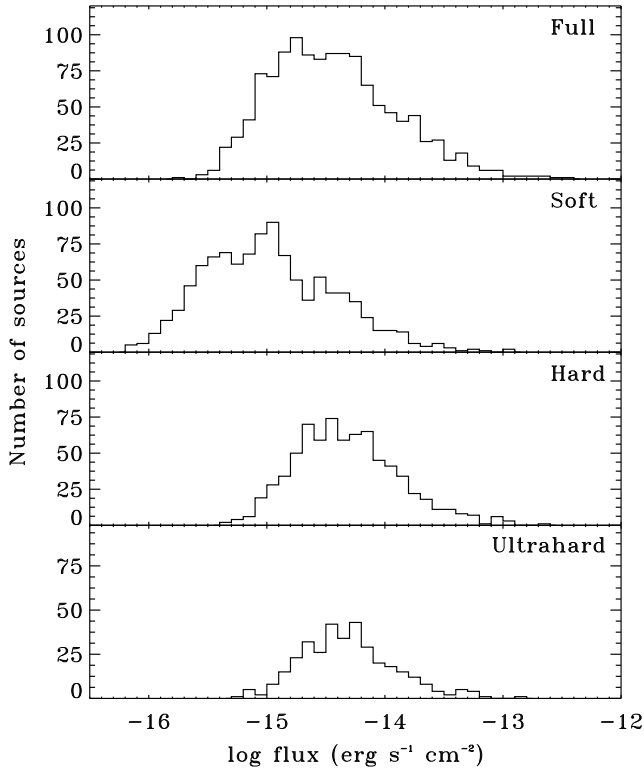}
}}
\caption{The distribution of fluxes for sources detected in the full, soft, hard, and ultrahards bands, using the Bayesian calculation of the fluxes.}
\label{fig:flux_hist}
\end{figure}

Using the sensitivity curves in Figure~\ref{fig:area_curve} we determine the limiting fluxes in each band, defined as the flux to which at least 1\% of the survey area is sensitive, to be $2.37\times10^{-16}$ (FB; 0.5--10~keV), $5.31\times10^{-17}$ (SB; 0.5--2~keV), $3.76\times10^{-16}$ (HB; 2--10~keV), and  $6.24\times10^{-16}$ erg~cm$^{-2}$~s$^{-1}$ (UB; 5--10~keV). The total area of the survey is 0.64 degree$^{2}$. Using classical sensitivity curves which do not correct for the Eddington bias, as in N05, gives limiting fluxes that are approximately a factor of 2 greater. 
Because the sensitivity varies dramatically across the images, we also give the flux limits corresponding to 50\% and 90\% completeness in Table~\ref{tab:limits}. 
Figure~\ref{fig:flux_hist} shows the distribution of X-ray flux in each of the four bands. The fluxes cover a range of over three dex, with $\sim$50\% of detected soft and hard band sources having fluxes of less than $9.7\times10^{-16}$  and $4.5\times10^{-15}$ erg~cm$^{-2}$~s$^{-1}$, respectively.

\begin{deluxetable}{lccc}
\tabletypesize{\small}
\tablewidth{0pt}
\tablecaption{AEGIS-X flux limits\label{tab:limits}}
\tablehead{
&
\multicolumn{3}{c}{Completeness limit\tablenotemark{a}} \\
\colhead{Band}&
\colhead{1\%\tablenotemark{b}} &
\colhead{50\%\tablenotemark{b}} &  
\colhead{90\%\tablenotemark{b}}  
}
\startdata
Full &$2.37$ & $13.03$ & $40.27$\\
Soft &$0.53$ & $3.35$ & $11.09$ \\
Hard &$3.76$ & $20.65$ & $62.37$ \\
Ultrahard &$6.24$ & $38.50$ & $124.45$ 
\enddata
\tablenotetext{a}{Flux to which 1, 50 and 90\% of the survey area is complete.}
\tablenotetext{b}{Fluxes are in units of $10^{-16}$ erg~cm$^{-2}$~s$^{-1}$.}
\end{deluxetable}

\subsection{Comparison with {\tt wavdetect}}
The standard procedure for source detection with \chandra\ ACIS data is the wavelet-based algorithm {\tt wavdetect} \citep{fr02}. In order to compare our detection method, and verify that it is correctly identifying significant sources, we also performed source detection using the CIAO {\tt wavdetect} algorithm in each of the 8 fields. Running {\tt wavdetect} with a detection threshold of $10^{-7}$, commonly used for ACIS-I surveys \citep[e.g.][]{al03,ya04,wa07}, and creating a band-merged catalog for the whole of the EGS we found the total number of independent sources detected was 1258, compared to 1325 using our method, and less sources were detected in every band  (Table~\ref{tab:det_methods}). 

Matching the two catalogs showed that there were 1208 sources in common. There were 59 sources that were only in the {\tt wavdetect} catalog and 122 sources that were only in the \ax\ catalog. To assess the likelihood of these sources being real we searched for optical and infrared (IR) counterparts in the DEEP2 and {\it Spitzer}/IRAC photometry catalogs covering the EGS (see \S\ref{sec:counterpart} for details of the optical and IR data and the method for identifying secure counterparts). Real X-ray sources are likely to have a higher counterpart identification rate. We find secure optical counterparts for 59\% of the \ax\--only sources, compared to an optical identification rate of 47\% for the {\tt wavdetect}-only sources. Of the 73  \ax\--only sources that are covered by the {\it Spitzer}/IRAC data 88\% have secure IR counterparts, compared to 40\% of the 43 {\tt wavdetect}-only sources covered by the data. Therefore while some fraction of the 59 X-ray sources that are in the {\tt wavdetect} catalog, but have been missed by our source detection, may indeed be real, we can be confident that the extra sources included in the \ax\ catalog are at least more likely to be real X-ray sources than those that were missed. 

\subsection{Comparison with N05 GWS analysis}

The three obsIDs that constitute EGS-8 were previously reduced and analyzed as part of the ``Groth-Westphal Strip'' survey and a point source catalog was presented in N05. In this more recent reduction and analysis of the data there are a number of differences in the source detection results, compared to the N05 results. As was stated in \S\ref{sec:reduction} this field was analyzed independently of the rest of the EGS data and therefore includes the same data as in our previous work.  In both analyzes 158 independent sources were detected in the field, however the IDs of all of the sources are not the same. There are 10 new \ax\ sources that were not detected in N05 and 9 sources in the N05 GWS catalog that are not included in the \ax\ catalog (Table~\ref{tab:gws}). One further GWS source (c13) was not detected in EGS-8 but was detected in EGS-6, which overlaps with the field, and is therefore included in the final source list. The high optical identification rate of the sources, as determined by \citet{ge06}, suggests that the majority of the 9 sources are probably real. However, in order to maintain a well controlled sample we have not included them in our final \ax\ catalog. 

There are several small differences in our reduction and analysis that could lead to the different source lists. The EGS-8 data used in this work has a longer total exposure ($\sim7$ks) and the PSFs were calculated using a different method and are on average 7\% larger than before. However the probable main reason that the sources are not detected in this analysis is the refinement of our source detection method from the method used in N05. The second pass of source detection described in \S\ref{sec:srcd} leads to 25\% higher background levels than in N05 and therefore faint sources with Poisson probabilities close to the detection threshold, such as those in Table~\ref{tab:gws}, will no longer pass the detection criteria. The Poisson probabilities of the 9 N05 sources using the new higher background are shown in Table~\ref{tab:gws}; several sources were just below our detection criteria for being included in the \ax\ catalog. 

\begin{deluxetable*}{cccccccc}
\tabletypesize{\small}
\tablewidth{0pt}
\tablecaption{GWS sources from N05 not included in AEGIS-X catalog\label{tab:gws}}
\tablehead{
\colhead{N05} & 
\colhead{RA J2000\tablenotemark{a}} &
\colhead{Dec J2000\tablenotemark{a}} &
\colhead{$p_{\rm min}$\tablenotemark{a,b}} &
\colhead{Det.\tablenotemark{a}} &
\colhead{$R_{AB}$\tablenotemark{c}} &
\colhead{Comments\tablenotemark{a}} & 
\colhead{$p_{\rm min}$\tablenotemark{b,d}}\\
\colhead{cat. no.} & 
\colhead{(J2000)} & 
\colhead{(J2000)} & &
\colhead{bands} &
\colhead{(mag)}
}
\startdata
c14  & 14 16 59.26 & +52 34 36.04 & $1.3\times10^{-6}$ & fh & 25.03 &         & $5.5\times10^{-6}$\\
c35  & 14 17 18.89 & +52 27 43.74 & $1.1\times10^{-6}$ & f  & 23.47 &chip gap & $4.1\times10^{-6}$\\
c45  & 14 17 25.28 & +52 35 12.08 & $4.0\times10^{-6}$ & f  &$>$26.0&         & $1.8\times10^{-4}$\\
c68  & 14 17 39.06 & +52 28 43.78 & $3.2\times10^{-7}$ & fs & 24.92 &chip gap & $2.3\times10^{-3}$\\
c69  & 14 17 39.31 & +52 28 50.16 & $6.3\times10^{-8}$ & fh & 23.26 &chip gap & $7.6\times10^{-4}$\\
c73  & 14 17 42.86 & +52 22 35.21 & $1.9\times10^{-8}$ & f  &$>$26.0&chip gap & $3.1\times10^{-3}$\\
c80  & 14 17 47.01 & +52 25 12.07 & $1.2\times10^{-6}$ & fh & 22.54 &         & $1.3\times10^{-5}$\\
c102 & 14 17 56.92 & +52 31 18.47 & $3.7\times10^{-6}$ & f  & 25.10 &         & $3.6\times10^{-5}$\\
c125 & 14 18 08.06 & +52 27 50.36 & $2.4\times10^{-6}$ & f  &$>$26.0&         & $2.7\times10^{-4}$
\enddata
\tablenotetext{a}{Values from N05.}
\tablenotetext{b}{Lowest false detection probability found for the four analysis bands.}
\tablenotetext{c}{Optical identification from \citet{ge06}.}
\tablenotetext{d}{Values from this work.}
\end{deluxetable*}

\subsection{Astrometric accuracy}
\label{sec:astrom}

\begin{deluxetable}{cccc}
\tabletypesize{\small}
\tablecaption{Absolute astrometric accuracy of sources in the AEGIS-X catalog \label{tab:astr_acc1}}
\tablewidth{0pt}
\tablehead{
 &
\multicolumn{3}{c}{Net full band counts} \\
\colhead{OAA} &
\colhead{$S \leqslant 50$} &
\colhead{$50 < S \leqslant 100$} & 
\colhead{$S > 100$} \\
\colhead{(arcmin)} &
\colhead{(arcsec)} &
\colhead{(arcsec)} &
\colhead{(arcsec)}
}
\startdata
0--4 & 0.56 & 0.36 & 0.14 \\
4--8 & 0.96 & 0.61 & 0.44 \\
$>8$  & 1.33 & 0.87 & 0.57
\enddata
\end{deluxetable}

The  astrometric accuracy  of  the \chandra\  positions was  estimated
using MARX to generate simulated fields.  Synthetic point sources were
added to the simulated images adopting  a single power-law  $\log N -
\log S$ distribution with parameters  tuned to reproduce the faint end
of the  2--10\,keV X-ray number  counts. The slope of  the differential
counts was  set to $\beta =  -1.58$ and the normalization  is fixed so
that there are 7000 sources per square degree brighter than $f_X(\rm 2
- 10 \, keV) = 10^{-16} \,  erg \, s^{-1}\, cm^{-2}$, i.e.  similar to the
observed density  of X-ray sources  \citep{ki07,ge08}.  
This  choice of parameters will over predict  the number of
bright  sources but  for the  purposes of  determining  the positional
error this is irrelevant. We adopt an X-ray spectrum with $\Gamma=1.4$
and a  total exposure time  of 200ks.  Simulated ACIS-I  event files
are constructed by randomly placing  within the {\it Chandra} field of
view point sources with fluxes in the  range $f_X(\rm 2 - 10\, keV ) =
5\times10^{-17} - 10^{-12}\, erg \, s^{-1} \, cm^{-2}$.  MARX does not
simulate  the Chandra  background.   This is  added  to the  simulated
images using the quiescent background event files produced by the ACIS
calibration  team using  blank sky  observations. A  total of  10 such
simulations were performed.  Each  simulated field was ran through the
entire source  detection procedure.  In each simulation,  there are on
average about 180 sources with detection probability $<4\times10^{-6}$
in  at least  one of  the four analysis bands.   The  detected source
positions were then compared to  the input positions of the sources in
the simulation. It should be noted that the astrometric accuracy determined this
way, using simple, single field images, accounts only for
statistical errors and does not take into account any systematic errors
that may be present using the more complex \ax\ data.

As expected, the positional accuracy  depends on both the OAA, and the
number of  source counts. However  the simulations showed only  a mild
dependence  on  source  counts   and  therefore  for  the  purpose  of
cross-band  matching we  chose radii  that depended  only on  OAA. The
estimated  $1\sigma$ positional  accuracies  are shown  in the  second
column of Table~\ref{tab:xmatch}, averaged over all source counts. 

\begin{figure}
\vspace{0.09in}
\hspace{-0.15in}
\epsscale{1.0}
{\scalebox{0.95}
{\includegraphics[angle=90]{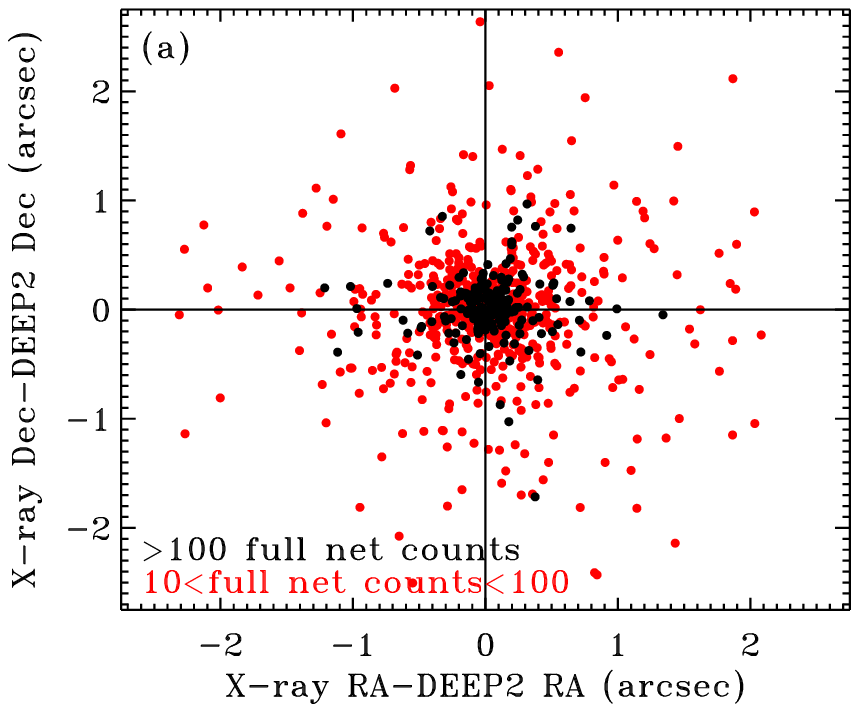}}
}
\vspace{0.1in}
\epsscale{1.0}
{\scalebox{0.95}
{\includegraphics[angle=90]{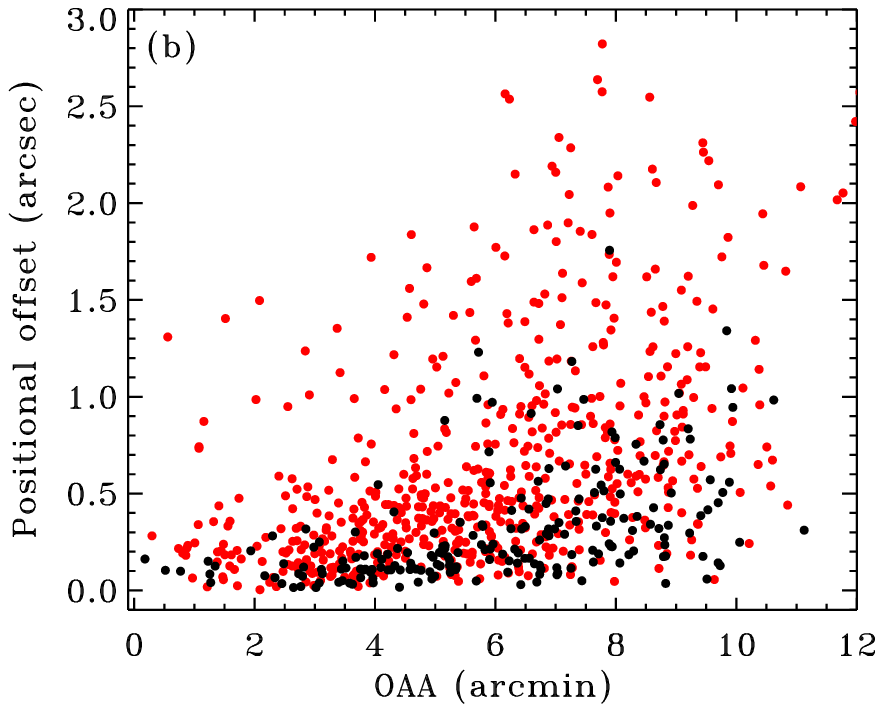}}
}
\caption{(a) The residual offset in RA and Declination between the matched \chandra\ and DEEP2 optical sources in the final AEGIS-X catalog, after the positional offsets were applied to each field (Table~\ref{tab:off}). (b) Total positional offset between the AEGIS-X and DEEP2 sources 
 vs. off-axis angle. The median and RMS positional offsets are $0\farcs40$ and $0\farcs79$, respectively. 
}
\label{fig:astrom}
\end{figure}

The absolute  astrometric accuracy of  the \chandra\ positions  in the
\ax\ catalog were  verified by matching the \ax\  sources to the DEEP2
optical data  from the {\it AEGIS} project  (see \S\ref{sec:optid} for
details), for which the  absolute positional uncertainty is $0\farcs2$
RMS (A. L. Coil, priv. communication).  
As  described above,  to combine the  catalogs from each  of the
eight  fields   we  first  registered  the  positions   to  match  the
astrometric  frame of  the DEEP2  catalog using  the offsets  given in
Table~\ref{tab:off}.   Matching  the   final  catalog  to  DEEP2,  and
considering only secure optical  counterparts, we find that the median
and RMS positional offsets of  868 matched sources are $0\farcs40$ and
$0\farcs79$, respectively.  The dependence of the astrometric accuracy
on   off-axis   angle   and   source   counts   are   illustrated   in
Figure~\ref{fig:astrom}.  We  determine  the $1\sigma$ astrometric  
accuracy  of
sources in the \ax\ catalog by breaking the sample down into nine bins
based on OAA and net counts (Table~\ref{tab:astr_acc1}).

\subsection{False source contamination}
The number of false sources in our catalog was assessed using the
quiescent background event files produced by the ACIS team based on
blank sky observations. These event files were used to simulate 20
200ks ACIS-I blank fields to match our data. The simulated images will
give an estimate of the number of statistical spurious sources
expected in 200ks data but will not account for possible systematic spurious
sources that may result from the more complex \ax\ data 
with its multiple overlapping observations. 
Running our full source detection procedure on the simulated images 
we find there to be 0.58 spurious sources detected per 200ks field per band, 
similar to the 0.5 sources per
band estimated by N05. Therefore we expect there to be a total of no more than 19
spurious sources in our catalog, equivalent to a contamination rate of
less than 1.5\%.

\subsection{Final \ax\ catalog}
The final \ax\ point source catalog is presented in Tables~\ref{tab:cat1} and \ref{tab:cat2}. Table~\ref{tab:cat1} gives the basic source properties such as source position, source counts, and details of the detection significance. Column (1) give the source ID and column (2) gives the field ID, which identifies the individual field in which the source was detected. Sources are listed in order of increasing RA. Column (3) gives the CXO object name. Columns (4) and (5) are the RA and Dec of the X-ray sources, respectively. These coordinates have been shifted to agree with the SDSS reference frame of the DEEP2 images (Table~\ref{tab:off}). Column (6) is the positional error in arcsec, as detailed above. The source OAA, in arcmin, is given in column (7). Columns (8)--(15) give the total source counts ($N$) and background counts ($B$) in the four analysis bands. Counts are given regardless of whether a source was detected in the band. Column (16) lists the bands in which the source is detected with Poisson probability $<4\times10^{-6}$, where the bands are full (f), soft (s), hard (h) and ultrahard (u). Column (17) gives the lowest false detection probability found for the four bands; probabilities lower than $10^{-8}$ are quoted in the table as $10^{-8}$.

Table~\ref{tab:cat2} lists the flux and HR information for the sources. Column (1) again gives the source ID. 
Columns (2)--(5) give the fluxes calculated using the Bayesian method described in \S\ref{sec:photom}, which corrects for the Eddington bias. The errors are 1$\sigma$ values and the limits are 1$\sigma$, or 68\%, upper limits.
In three cases (egs\_0367, egs\_0457, and egs\_0633) sources that were significantly detected in a band had best
 estimate fluxes of zero, using the Bayesian method. This discrepancy arises due to the different EEFs used for the photometry (90\% EEF) and for the detection (70\% EEF). In these cases we use source and background counts corresponding to the 70\% EEF area to calculate the fluxes in the detected bands. Columns (6)--(9) give the observed frame fluxes in the four bands, where the fluxes and errors have been calculated using the classical method. The errors are 1$\sigma$ values, however here the limits are 99\% upper limits. All the fluxes listed are in units of $10^{-15}$ erg cm$^{-2}$ s$^{-1}$ and have not been corrected for Galactic or intrinsic source absorption. To correct for Galactic absorption the full and soft band fluxes should be increased by 2\% and 4.2\%, respectively. The corrections to the hard and ultrahard fluxes are negligable. Columns (10) and (11) give the Bayesian and classical HRs, respectively. The errors for the Bayesian HRs are 1$\sigma$. For sources only detected in the full band, and with upper limits in both hard and soft bands, the clasical hardness ratio cannot be determined and is set to $-99$. Column (12) is a flag for the quality of the photometry. A flag of `1' indicates the presence of a nearby source that may be contaminating the photometry. A flag of `2' indicates that another source was detected with the 90\% EEF and that the photometry is likely heavily contaminated and the source position uncertain (\S\ref{sec:photom}). All other sources have a flag of `0'.

The \ax\ source catalog is publicly available in FITS table format at \\\url{http://astro.imperial.ac.uk/research/aegis/cats.shtml}. The FITS version of the catalog includes more information about the sources, such as the Poisson probability, counts, PSF size, and exposure value in each of the bands. Postage stamp images of the sources are also available.

\section{Optical and IR identifications}
\label{sec:counterpart}

\begin{figure}
\hspace{-0.050in}
\epsscale{1.0}
{\scalebox{1.0}
{\includegraphics{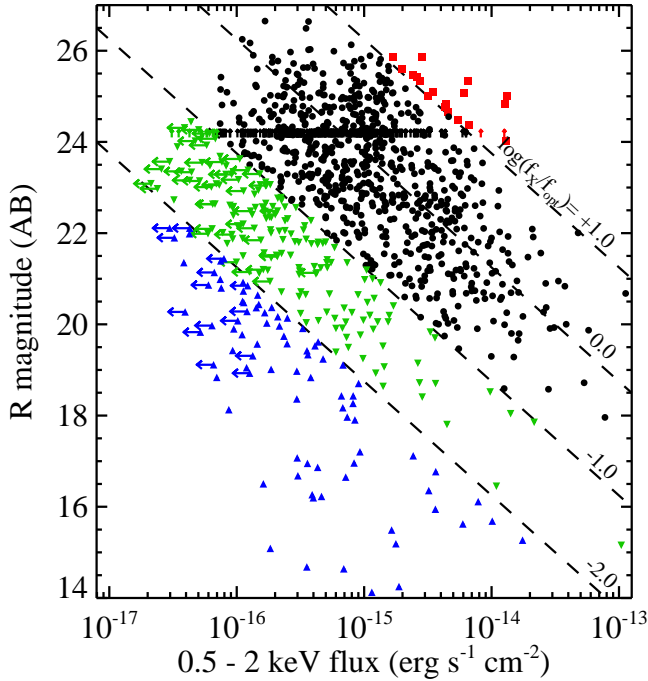}}
}
\caption{0.5--2 keV flux versus $R$-magnitude for sources with DEEP2 or CFHTLS counterparts.  For X-ray sources with DEEP2 counterparts we plot the DEEP2 magnitudes, for sources not detected in DEEP2 but detected in the CFHTLS then we use CFHTLS $r\arcmin_{AB}$ converted to $R_{AB}$ (as described in \S\ref{sec:optid}), otherwise we plot the DEEP2 upper limit. Lines of constant X-ray-to-optical flux ratio are plotted, according to the relation of \citet{ho01}. The symbol shapes and colors are defined by the positions on the $f_{\mathrm{X}}/f_{\mathrm{opt}}$ plot as follows: 
(red squares) log$(f_{\mathrm{X}}/f_{\mathrm{opt}})>1$, (black circles) $-1<$log$(f_{\mathrm{X}}/f_{\mathrm{opt}})<1$, (green inverted triangles)  $-2<$log$(f_{\mathrm{X}}/f_{\mathrm{opt}})<-1$, and (blue triangles) log$(f_{\mathrm{X}}/f_{\mathrm{opt}})<-2$.}
\label{fig:fxfopt}
\end{figure}

We identify secure optical and  IR counterparts to the \chandra\ X-ray
sources  using the  maximum likelihood  (LR)  method \citep{ci03,ci05,br07}.  
This is the ratio between
the  probability that a  source, at  a given  distance from  the X-ray
position and with a given optical  or IR magnitude, is the true ID and
the probability that this source is a spurious alignment. The LR ratio
takes into  account the X-ray, optical or  IR positional uncertainty,
the background  density of optical (or  IR) galaxies and  the a priori
probability of a counterpart with a given magnitude.

For the positional  accuracy of the X-ray sources  we adopt a Gaussian
distribution  with  standard  deviation  estimated as  a  function  of
off-axis angle and total number  of counts from the  MARX simulations
described in  \S\ref{sec:astrom}. 
The optical identification  is a  two pass
procedure.  In the first pass we estimate the a priori probability that
an X-ray  source has  a counterpart of  given magnitude in  the input
catalog  (optical or  IR). We  use  a fixed  search radius  of
2\arcsec~to identify the X-ray sources with counterparts in the input
catalog.  A  total of  100 mock catalogs  are then  constructed by
randomising  the  source positions  of  the  input  catalog and  the
cross-matching  is   repeated.   The  magnitude   distribution  of  the
counterparts in  the mock  and the real  catalogs are  subtracted to
determine the  magnitude distribution of the  real associations.  This
is then  used as a prior in the  LR ratio estimation to  account for the
distribution  of X-ray sources  in magnitudes  and the  total expected
number of  true associations.  In the  second pass a  search radius of
4$\arcsec$ (equivalent to the 3$\sigma$ positional uncertainty at large OAA)
is  adopted to identify all possible  counterparts to X-ray
sources and to estimate the LR of each one. Where more than one counterpart 
is found within the search radius we identify the one with the highest LR 
to be the most likely counterpart.

\subsection{Optical identifications}
\label{sec:optid}

There is extensive optical photometric coverage of the AEGIS field from a number of telescopes, including SDSS, KPNO, Subaru and MMT (further details can be found in \citealt{da07}). Here we consider two sets of optical photometric data, both taken at the Canada-France-Hawaii Telescope (CFHT),  as our primary matching datasets. The DEEP2 photometric imaging in the $B$, $R$, and $I$ bands is described in \citet{co04}. These data almost the entire \chandra\ AEGIS field and are complete to a limiting magnitude of $R_{AB}=24.1$. The astrometric accuracy of the photometric catalog is estimated to be $0\farcs2$.

We also consider deeper CFHT data that were obtained for a 1~deg$^{2}$ area as part of the CFHT Legacy survey\footnote{see http://www.cfht.hawaii.edu/Science/CFHLS/ and http://www.ast.obs-mip.fr/article204.html.} (CFHTLS). The CFHTLS includes data from five filters: $u^{\ast}$, $g\arcmin$, $r\arcmin$, $i\arcmin$, and $z\arcmin$, and are complete to $i\arcmin_{AB}=27.0$. The overlap between the CFHTLS imaging area and the \chandra\ area is 0.33~deg$^{2}$, considerably smaller than the overlap with the DEEP2 optical imaging described above. The CFHTLS  catalogs we  are using are from the T0003 release, in which the astrometry was calculated  using the  USNO-B1 catalog. The astrometric uncertainty of the CFHTLS catalog is estimated to be about $0\farcs3$. Details about the data and catalog are given the TERAPIX pages relevant to that release: \url{http://terapix.iap.fr/article.php?id\_article=556}.

Using the maximum likelihood method we search for optical counterparts to the \ax\ sources in the DEEP2 survey using the DEEP2 $R$-selected catalog to estimate the surface density of background optical sources, $R$ being the deepest band in the DEEP2 survey. The $i\arcmin$ band is the deepest band in the CFHTLS survey so we use the $i\arcmin$-selected catalog to estimate the surface density of background optical sources for the CFHTLS. For both the DEEP2 and CFHTLS surveys we consider counterparts with LR$>$0.5 as secure, giving a $\sim$6\% contamination rate in both cases. 

Of the 1317 \ax\ sources that are covered by the DEEP2 survey, which is complete to $R_{AB}=24.1$, we find that 897 have secure $R$ band counterparts an optical ID rate of 68.1\%. 
The deeper CFHTLS data, complete to $i\arcmin_{AB}=27.0$, overlaps with 703 \ax\ sources, and we find that 578 (82.2\%) of them have secure $i\arcmin$ band selected optical counterparts. Taking into account both catalogs, 1013 of the 1325 \ax\ sources have optical counterparts, with $13.95\leqslant R_{AB}\leqslant 26.65$. At the limit to which all of the optical coverage is complete, $R_{AB}=24.1$, we find an optical counterpart rate of 57\%. The optical ID rates found for the \ax\ sources are similar to those found for other deep \chandra\ surveys. For instance, in the deeper CDF-N \citet{ba03} found optical counterparts for 85\% of the sources to $R\leqslant 26.4$, with an estimated 15\% contamination rate. Considering only $R\leqslant 24.0$ sources, they found an optical ID rate of 55\%, in agreement with the \ax\ findings. 

Table~\ref{tab:optid} gives the results of the optical counterpart analysis. The table lists the $R$ magnitude for every source with a secure DEEP2 counterpart, along with the offset between the X-ray and DEEP2 source ($\delta_{opt-X}$), the LR, and the DEEP2 counterpart ID number. The $i\arcmin$ magnitude, offset between the  X-ray and CFHTLS source, LR, and counterpart ID number are also given for every secure CFHLTS counterpart. 

Figure~\ref{fig:fxfopt} shows soft X-ray flux versus $R$ magnitude for the sample, for all the sources with secure optical counterparts.  For X-ray sources with DEEP2 counterparts we use the DEEP2 $R$ magnitude. For sources not detected in the DEEP2 survey, but securely detected in the CFHTLS then we convert the CFHTLS $r\arcmin_{AB}$ magnitudes to $R_{AB}$ using 
\begin{equation}
        R=r\arcmin - 0.0576 -0.3718[(r\arcmin-i\arcmin)-0.2589]
\end{equation}
\citep{bl07}. This transformation is correct for the \citet{be98} $R$ filter which differs slightly from the CFHT filter used in the DEEP2 survey, but should be correct to within 0.1 magnitudes or less. Upper limits are plotted, where appropriate.  The ratio of X-ray to optical flux is a useful diagnostic for the determining the nature of X-ray sources. Over several decades in flux classic AGN, both narrow and broad line, exhibit X-ray to optical flux ratios of $-1<$log$(f_{X}/f_{opt})<+1$ \citep[e.g.][]{sc98,ak00,al01}, while at lower X-ray to optical flux ratios low luminosity AGN, starburst galaxies, and normal galaxies emerge \citep[e.g.][]{al01,ba02}. The sources in the \ax\ catalog cover over four orders of magnitudes in $f_{X}$-to-$f_{opt}$ ratio. As expected, the majority of the \ax\ sources populate the AGN region of the diagram but a significant number of sources fall within the regions normally populated with starburst or composite star-forming/AGN galaxies ($-2<$log$(f_{X}/f_{opt})<-1$; e.g. \citealt{bau02,al02}) and normal galaxies (log$(f_{X}/f_{opt})<-2$; \citealt{ho03}).

\subsection{IRAC identifications}

\begin{figure}
\hspace{-0.05in}
\epsscale{1.0}
{\scalebox{1.0}
{\includegraphics{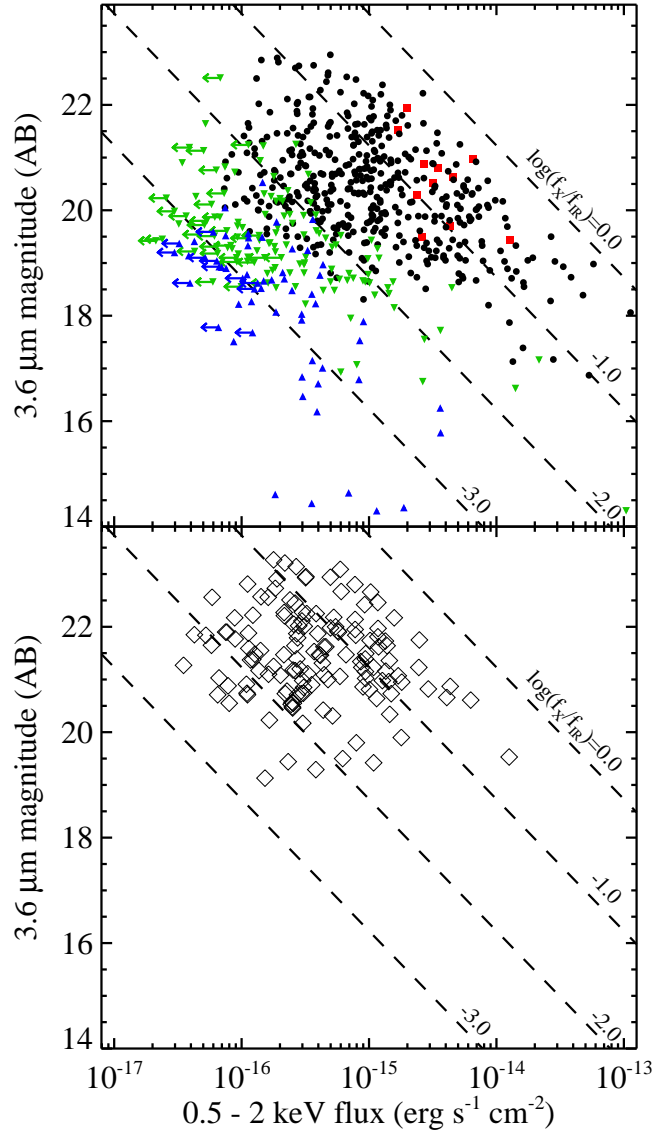}}
}
\caption{0.5--2 keV flux versus IRAC 3.6$\micron$ magnitudes for sources with secure IRAC counterparts.  Lines of constant X-ray-to-IR flux ratio are plotted, using the 3.6$\micron$ zero-points given by \citet{re05}. (Top) Sources with secure optical counterparts: symbol shapes and colors are the same as Fig~\ref{fig:fxfopt}. (Bottom) Sources without secure optical counterparts. A few bright, $m_{3.6\micron} < 19.0$, sources that do not have secure optical counterparts listed in Table~\ref{tab:optid} have been removed from this plot as visual inspection revealed clear bright, optical counterparts with mismatching optical positions, such as saturated stars. Such sources are commented in Table~\ref{tab:optid}.}
\label{fig:fxfir}
\end{figure}

The majority of the AEGIS field is covered by sensitive {\it Spitzer}/IRAC data, described in detail by \citet{ba06} and \citet{ba08}. Briefly, these data cover the whole length of the strip but with a width of only 10$\arcmin$. Here we consider only the 3.6$\micron$-selected catalog of \citet{ba08}, which has a 50\% point source completeness limit of  0.9 $\mu$Jy at 3.6 $\mu m$ (or $m_{3.6\micron}$(AB)=23.8). The astrometric uncertainty of the IRAC catalog is estimated to be $0\farcs35$. A total of 882 \chandra\ sources are covered by the {\it Spitzer} region. Using the maximum likelihood method, and again considering only matches with LR$>$0.5, we find that 830 (94.1\%) have secure 3.6$\micron$ counterparts. This cut on LR gives a $\sim$1\% spurious counterpart rate. The 94\% IRAC ID rate found here is consistent with that found in the 250ks Extended \chandra\ Deep Field-South; \citet{ca08} found IRAC counterparts for 90\% of the X-ray sources to an IRAC flux limit of 0.63 $\mu$Jy at 3.6 $\mu m$.

Table~\ref{tab:optid} lists the 3.6$\micron$ magnitude for each source with a secure IRAC counterpart, along with the offset between the X-ray and IRAC source ($\delta_{IR-X}$), the LR, and the IRAC ID from \citet{ba08}. Figure~\ref{fig:fxfir} shows soft X-ray flux versus 3.6$\micron$ magnitude for the sample, for the sources with an IRAC counterpart. Galaxies identified as classic AGN in Fig~\ref{fig:fxfopt} (black circles and red squares) occupy the region of the plot defined by $-2<$log$(f_{X}/f_{IR})<0$. The sources in the \ax\ catalog cover over three orders of magnitudes in $f_{X}$-to-$f_{IR}$ ratio.  The \ax\ sources without an optical counterpart are on average fainter at 3.6$\micron$ and exhibit a smaller range in $f_{X}$-to-$f_{IR}$ ratio than those sources with an optical counterpart. 

\section {Conclusions}
\label{sec:conc}

We have presented a catalog of X-ray point sources from the 0.67~deg$^2$ \ax\ survey. Using a detection threshold of Poisson probability $<4\times10^{-6}$ we detected 1325 independent point sources down to on-axis flux limits of $2.37\times10^{-16}$ (0.5--10~keV), $5.31\times10^{-17}$ (0.5--2~keV) and $3.76\times10^{-16}$ erg~cm$^{-2}$~s$^{-1}$ (2--10~keV). For each source we determine the X-ray flux in four bands (full, soft, hard, and ultrahard) using a Bayesian method that corrects for the Eddington bias, in addition to the standard flux calculation method. Optical and IR counterparts to the X-ray sources were identified from the DEEP2, CFHTLS, and {\it Spitzer}/IRAC surveys of the EGS, and the results presented. We find that 76\% and 94\% of the \ax\ sources have secure optical and IR counterparts, respectively. 

A number of the analysis methods used in this work were specifically developed for \chandra\ surveys such as this one. The source catalog was produced using our own source detection procedure that is described in detail. Simulations verified the validity of this detection procedure and predict a very small, $<1.5$\%, contamination rate from spurious sources. The photometry method employed corrects for the Eddington bias. A method for calculating accurate PSFs for multi-observation datasets, such as the AEGIS survey, is also presented. Sensitivity maps and sensitivity curves were made following the new method of \citet{ge08}, which efficiently and correctly accounts for the observational biases that affect the probability of detecting a source at a given flux.  

Finally, all of the data products described in this paper, including reduced event files, images, exposure maps, sensitivity maps, sensitivity curves, and PSF tables (which can be used for any \chandra\ ACIS-I data), are available via the public website \\\url{http://astro.imperial.ac.uk/research/aegis/}. The \ax\ catalog and the optical and IR counterparts catalog are also made available.



\acknowledgments
We thank the referee for helpful comments that improved clarity of the paper. We thank those who have built and operate the \chandra\ X-ray observatory so successfully. We acknowledge financial support from the Leverhulme trust (KN), PPARC/STFC (ESL, JA), Marie-Curie fellowship grant MEIF-CT-2005-025108 (AG), and \chandra\ grant GO5-6141A (DCK). 
Facilities: \facility{CXO(ACIS)}, \facility{CFHT}, \facility{Spitzer(IRAC)}.

\clearpage

\begin{landscape}
\begin{deluxetable}{lllcccccccccccccc}
\tabletypesize{\tiny}
\tablecaption{\chandra\ \ax\ source catalog: basic source properties\label{tab:cat1}}
\tablewidth{0pt}
\tablehead{
\colhead{} &      
\colhead{} &      
\colhead{} &      
\colhead{RA} &            
\colhead{Dec} &           
&         
&         
\multicolumn{2}{c}{FB cts} & 
\multicolumn{2}{c}{SB cts} & 
\multicolumn{2}{c}{HB cts} & 
\multicolumn{2}{c}{UB cts} & 
\colhead{Detection} & 
\\        
\colhead{ID} &            
\colhead{Field ID} &      
\colhead{CXOEGS} &        
\colhead{(J2000)} &       
\colhead{(J2000)} &       
\colhead{Pos. err} &      
\colhead{OAA} &   
\colhead{N} &     
\colhead{B} &     
\colhead{N} &     
\colhead{B} &     
\colhead{N} &     
\colhead{B} &     
\colhead{N} &     
\colhead{B} &     
\colhead{bands} & 
\colhead{$p_{min}$} \\  
\colhead{(1)} &
\colhead{(2)} &
\colhead{(3)} &
\colhead{(4)} &
\colhead{(5)} &
\colhead{(6)} &
\colhead{(7)} &
\colhead{(8)} &
\colhead{(9)} &
\colhead{(10)} &
\colhead{(11)} &
\colhead{(12)} &
\colhead{(13)} &
\colhead{(14)} &
\colhead{(15)} &
\colhead{(16)} &
\colhead{(17)} 
}
\startdata
egs\_0001 & egs7\_198 & J141410.9+520503 & 213.545293 & 52.084321 & 1.33 & 11.68 & 28 & 12.6 & 14 &  3.0 & 15 &  9.9 &  8 &  6.6 & s   & $10^{-6.75}$ \\
egs\_0002 & egs7\_129 & J141418.2+520856 & 213.575718 & 52.149002 & 1.33 & 10.11 & 49 & 21.0 & 24 &  5.4 & 26 & 17.2 & 13 & 10.6 & fs  & $10^{-8.00}$ \\
egs\_0003 & egs7\_111 & J141419.0+520404 & 213.579288 & 52.067991 & 1.33 & 10.82 & 59 & 12.0 & 31 &  2.9 & 29 &  9.2 & 11 &  5.6 & fsh & $10^{-8.00}$ \\
egs\_0004 & egs7\_128 & J141423.0+520809 & 213.595988 & 52.135866 & 1.33 & 9.35  & 49 & 24.4 & 21 &  6.1 & 32 & 19.2 & 16 & 12.3 & fs  & $10^{-6.32}$ \\
egs\_0005 & egs7\_193 & J141423.2+520708 & 213.596674 & 52.118981 & 1.33 & 9.39  & 38 & 23.5 & 16 &  5.4 & 22 & 18.9 & 14 & 12.7 & s   & $10^{-5.41}$ \\
egs\_0006 & egs7\_183 & J141424.0+520137 & 213.599946 & 52.027186 & 1.33 & 11.36 & 42 & 16.0 & 18 &  4.4 & 23 & 11.8 & 13 &  7.3 & f   & $10^{-8.00}$ \\
egs\_0007 & egs7\_034 & J141426.5+521136 & 213.610293 & 52.193375 & 1.33 & 9.41  & 60 & 18.7 & 25 &  4.4 & 35 & 15.4 & 10 & 10.1 & fsh & $10^{-8.00}$ \\
egs\_0008 & egs7\_110 & J141428.2+520346 & 213.617692 & 52.062882 & 1.33 & 9.67  & 58 & 16.6 & 39 &  4.3 & 19 & 13.1 &  5 &  8.5 & fs  & $10^{-8.00}$ \\
egs\_0009 & egs7\_122 & J141429.8+520659 & 213.624364 & 52.116556 & 0.87 & 8.40  & 72 & 21.6 & 37 &  5.2 & 35 & 17.8 & 15 & 11.8 & fs  & $10^{-8.00}$ \\
egs\_0010 & egs7\_127 & J141431.1+520357 & 213.629634 & 52.065995 & 1.33 & 9.19  & 53 & 22.6 & 31 &  5.7 & 25 & 17.6 & 12 & 11.3 & fs  & $10^{-8.00}$   
\enddata
\tablecomments{Table \ref{tab:cat1} is published in  its entirety in the electronic edition of the journal.}
\end{deluxetable}

\begin{deluxetable}{lllcccccccccc}
\tabletypesize{\scriptsize}
\tablecaption{\chandra\ \ax\ source catalog: source fluxes and HRs\label{tab:cat2}}
\tablewidth{0pt}
\tablehead{
\colhead{} &      
\multicolumn{4}{c}{Bayesian flux} & 
& 
\multicolumn{4}{c}{Classical flux} & 
\colhead{Bayesian}  &      
\colhead{Classical} &      
\colhead{Phot.}\\      
\cline{2-5}
\cline{7-10}
\colhead{ID} &            
\colhead{$f_{0.5-10}$}&   
\colhead{$f_{0.5-2}$} &   
\colhead{$f_{2-10}$}  &   
\colhead{$f_{5-10}$}  &   
&
\colhead{$f_{0.5-10}$}&   
\colhead{$f_{0.5-2}$} &   
\colhead{$f_{2-10}$}  &   
\colhead{$f_{5-10}$}  &   
\colhead{HR}          &   
\colhead{HR}          &   
\colhead{flag}        \\  
\colhead{(1)} &
\colhead{(2)} &
\colhead{(3)} &
\colhead{(4)} &
\colhead{(5)} & &
\colhead{(6)} &
\colhead{(7)} &
\colhead{(8)} &
\colhead{(9)} &
\colhead{(10)} &
\colhead{(11)} &
\colhead{(12)} 
}
\startdata
egs\_0001 & $5.16^{+2.35}_{-2.13}$ & $1.66^{+0.73}_{-0.60}$ & $<3.74$ & $<3.93$                && $6.42^{+2.65}_{-2.19}$ & $2.03^{+0.89}_{-0.68}$ & $<6.70$ & $<9.52$                  & $-0.37^{+0.30}_{-0.32}$ & -1.00 & 0 \\
egs\_0002 & $4.80^{+1.39}_{-1.27}$ & $1.39^{+0.43}_{-0.37}$ & $<2.79$ & $<2.31$                && $5.32^{+1.53}_{-1.33}$ & $1.55^{+0.50}_{-0.41}$ & $<3.74$ & $<5.51$                  & $-0.38^{+0.28}_{-0.24}$ & -1.00 & 0 \\
egs\_0003 & $17.68^{+3.11}_{-2.85}$ & $4.57^{+0.99}_{-0.88}$& $12.30^{+3.29}_{-3.66}$& $<5.79$ && $18.42^{+3.42}_{-3.00}$ & $4.85^{+1.15}_{-0.96}$ & $13.99^{+4.55}_{-3.77}$ & $<8.90$ & $-0.22^{+0.18}_{-0.14}$ & -0.22 & 0 \\
egs\_0004 & $2.73^{+0.95}_{-0.87}$ & $0.71^{+0.27}_{-0.23}$ & $1.72^{+1.08}_{-1.72}$ & $<1.82$ && $3.15^{+1.03}_{-0.89}$ & $0.83^{+0.32}_{-0.25}$ & $2.99^{+1.57}_{-1.31}$ & $<3.81$   & $-0.13^{+0.28}_{-0.25}$ & -0.12 & 0 \\
egs\_0005 & $1.12^{+0.57}_{-0.82}$ & $0.44^{+0.23}_{-0.20}$ & $<1.15$ & $<1.49$                && $1.79^{+0.90}_{-0.76}$ & $0.57^{+0.28}_{-0.21}$ & $<2.73$ & $<3.92$                  & $-0.40^{+0.30}_{-0.40}$ & -1.00 & 0 \\
egs\_0006 & $9.68^{+2.80}_{-2.53}$ & $2.10^{+0.82}_{-0.69}$ & $5.34^{+2.80}_{-3.70}$ & $<6.31$ && $10.72^{+3.11}_{-2.66}$ & $2.48^{+0.97}_{-0.77}$ & $8.23^{+4.31}_{-3.49}$ & $<9.69$  & $-0.16^{+0.27}_{-0.25}$ & -0.15 & 0 \\
egs\_0007 & $7.66^{+1.57}_{-1.44}$ & $1.60^{+0.45}_{-0.39}$ & $5.94^{+2.23}_{-2.02}$ & $<1.95$ && $8.09^{+1.72}_{-1.51}$ & $1.76^{+0.52}_{-0.42}$ & $6.97^{+2.49}_{-2.10}$ & $<5.33$   & $-0.07^{+0.20}_{-0.18}$ & -0.07 & 0 \\
egs\_0008 & $7.93^{+1.59}_{-1.46}$ & $2.92^{+0.57}_{-0.51}$ & $<2.12$ & $<1.59$                && $8.36^{+1.75}_{-1.53}$ & $3.07^{+0.65}_{-0.55}$ & $<3.62$ & $<5.28$                  & $-0.69^{+0.14}_{-0.18}$ & -1.00 & 0 \\
egs\_0009 & $4.55^{+0.83}_{-0.76}$ & $1.23^{+0.26}_{-0.23}$ & $2.39^{+1.10}_{-1.02}$ & $<1.27$ && $4.76^{+0.90}_{-0.80}$ & $1.30^{+0.29}_{-0.25}$ & $2.99^{+1.22}_{-1.03}$ & $<2.67$   & $-0.33^{+0.18}_{-0.17}$ & -0.33 & 0 \\
egs\_0010 & $3.15^{+0.86}_{-0.79}$ & $1.15^{+0.29}_{-0.25}$ & $<1.47$ & $<1.26$                && $3.46^{+0.95}_{-0.82}$ & $1.25^{+0.33}_{-0.27}$ & $<2.39$ & $<3.42$                  & $-0.54^{+0.21}_{-0.23}$ & -1.00 & 0     
\enddata
\tablecomments{Table \ref{tab:cat2} is published in its entirety in the electronic edition of the journal.}
\end{deluxetable}
\clearpage
\end{landscape}

\clearpage
\begin{landscape}
\begin{deluxetable}{lcccccccccccccc}
\tabletypesize{\scriptsize}
\tablecaption{Optical and IR counterparts to the \ax\ sources\label{tab:optid}}
\tablewidth{0pt}
\tablehead{
&
\multicolumn{4}{c}{DEEP2} & & 
\multicolumn{4}{c}{CFHTLS} & &
\multicolumn{4}{c}{IRAC} \\
\cline{2-5}
\cline{7-10}
\cline{12-15}
\colhead{ID}  &
\colhead{$m_{\mathrm {R}}$\tablenotemark{a}} &
\colhead{$\delta_{opt-X}$\tablenotemark{b}} &
\colhead{LR$_{opt}$} &   
\colhead{ID} & &
\colhead{$m_{\mathrm {i\arcmin}}$\tablenotemark{a}} &
\colhead{$\delta_{opt-X}$\tablenotemark{b}} &
\colhead{LR$_{opt}$} &   
\colhead{ID} & &
\colhead{$m_{\mathrm {3.6\micron}}$\tablenotemark{a}} &
\colhead{$\delta_{IR-X}$\tablenotemark{c}} & 
\colhead{LR$_{IR}$}  & 
\colhead{ID\tablenotemark{d}}\\
& 
\colhead{(AB)} &
\colhead{($\arcsec$)} &
& & &
\colhead{(AB)} &
\colhead{($\arcsec$)} & 
& & &
\colhead{(AB)} &
\colhead{($\arcsec$)}  
}
\startdata
egs\_0001 & 22.84 & 2.02 & 1.55 & 11021738 && \ldots  & \ldots  & \ldots  & \ldots   && \ldots & \ldots & \ldots & \ldots \\
egs\_0002 & 23.53 & 1.04 & 4.15 & 11028286 && \ldots  & \ldots  & \ldots  & \ldots   && \ldots & \ldots & \ldots & \ldots \\
egs\_0003 & 21.11 & 1.65 & 7.14 & 11015296 && \ldots  & \ldots  & \ldots  & \ldots   && \ldots & \ldots & \ldots & \ldots \\
egs\_0004 & 22.00 & 1.49 & 7.68 & 11028607 && \ldots  & \ldots  & \ldots  & \ldots   && \ldots & \ldots & \ldots & \ldots \\
egs\_0005 & 21.80 & 1.15 & 13.01 & 11021753 && \ldots  & \ldots  & \ldots  & \ldots  & & \ldots & \ldots & \ldots & \ldots \\
egs\_0006 & $>$24.1 & \ldots   & \ldots  & \ldots  && \ldots  & \ldots  & \ldots  & \ldots   &&  $>$23.8 & \ldots & \ldots & \ldots \\
egs\_0007 & 22.30 & 0.64 & 31.25 & 11035367 && \ldots  & \ldots  & \ldots  & \ldots  & & \ldots & \ldots & \ldots & \ldots \\
egs\_0008 & $>$24.1 & \ldots   & \ldots  & \ldots  && \ldots  & \ldots  & \ldots  & \ldots   && 20.82 & 1.31 & 9.46 & J141428.13+520347.1 \\
egs\_0009 & 23.94 & 0.70 & 10.04 & 11021149 && \ldots  & \ldots  & \ldots  & \ldots   && \ldots & \ldots & \ldots & \ldots \\
egs\_0010 & 22.66 & 1.26 & 6.69 & 11014844 && \ldots  & \ldots  & \ldots  & \ldots   && 20.25 & 1.09 & 25.36 & J141431.06+520358.6
\enddata
\tablecomments{Table \ref{tab:optid} is published in  its entirety in the electronic edition of the journal.}
\tablenotetext{a}{If optical or IR coverage exists but no secure counterpart was detected then an upper limit for the magnitude is given. X-ray sources with no optical or IR coverage are denoted as `\ldots'.}
\tablenotetext{b}{The offset between the X-ray and optical position.}
\tablenotetext{c}{The offset between the X-ray and IRAC position.}
\tablenotetext{d}{Truncated IRAC object ID number from \citet{ba08}. Object IDs should be preceded by EGSIRAC.}
\end{deluxetable}
\clearpage

\end{landscape}

\end{document}